\documentclass[12pt,preprint]{aastex}

\slugcomment{Accepted to ApJ}

\shorttitle{SLIP-SQUASHING FACTORS AS A MEASURE OF 3D REC}
\shortauthors{TITOV}

\newcommand{\hl}{}
\newcommand{\bm}[1]{ \mbox{\boldmath{$#1$}} }

\newcommand{\RFfp }{\underline{\mathrm{RF} }}
\newcommand{\QSSfp}{\underline{\mathrm{QSS}}}
\newcommand{\tr}{\mathop{\mathrm{tr}}\nolimits}
\newcommand{\ad}{* \hspace{-0.6pt} *}
\newcommand{\aaa}{\raisebox{+0.2em}{$\scriptstyle *\hspace{-0.6pt}*$}
  \hspace{-0.52em} \raisebox{-0.07em}{$\scriptstyle *$}}
\newcommand{\aaas}{\raisebox{+0.15em}[1pt]{$\scriptscriptstyle *\hspace{-0.7pt}*$}
  \hspace{-0.46em} \raisebox{-0.05em}{$\scriptscriptstyle *$}}
\newcommand{\aaaa}{\raisebox{+0.25em}{$\scriptstyle *\hspace{-0.6pt}*$}
  \hspace{-0.67em} \raisebox{-0.1em}{$\scriptstyle *\hspace{-0.6pt}*$}}
 
\begin{document}

\title{SLIP-SQUASHING FACTORS AS A MEASURE OF THREE-DIMENSIONAL MAGNETIC RECONNECTION} 


\author{V. S. Titov\altaffilmark{1}}
\affil{Science Applications International Corporation, 10260 Campus
Point Drive, San Diego, California 92121-1578} 
\email{titovv@predsci.com}

\and

\author{T. G. Forbes} 

\affil{University of New Hampshire, EOS Institute 8 College Road, Durham, NH 03824}
\email{terry.forbes@unh.edu}

\and

\author{E. R. Priest} 

\affil{School of Mathematics and Statistics, University of St. Andrews, Fife KY16 9SS, Scotland, UK}
\email{eric@mcs.st-and.ac.uk}

\and

\author{Z. Miki\'{c}\altaffilmark{1}, J.~A. Linker\altaffilmark{1}}
\affil{Science Applications International Corporation, 10260 Campus
Point Drive, San Diego, California 92121-1578} 
\email{mikicz@predsci.com, linkerj@predsci.com}

\altaffiltext{1}{In transition to Predictive Science, Inc., 9990 Mesa Rim Road, Suite 170, San Diego, CA 92121-2910}

\begin{abstract}
A general method for describing magnetic reconnection in arbitrary three-dimensional magnetic configurations is proposed.
The method is based on the field-line mapping technique previously used only for the analysis of magnetic structure at a given time.
This technique is extended here so as to analyze the evolution of magnetic structure.
Such a generalization is made with the help of new dimensionless quantities called ``slip-squashing factors".
Their large values define the surfaces that border the reconnected or to-be-reconnected magnetic flux tubes for a given period of time during the magnetic evolution.
The proposed method is universal, since it assumes only that the time sequence of evolving magnetic field and the tangential boundary flows are known.
The application of the method is illustrated for simple examples, one of which was considered previously by Hesse and coworkers in the framework of the general magnetic reconnection theory.
The examples help us to compare these two approaches; they reveal also that, just as for magnetic null points, hyperbolic and cusp minimum points of a magnetic field may serve as favorable sites for magnetic reconnection.
The new method admits a straightforward numerical implementation and provides a powerful tool for the diagnostics of magnetic reconnection in numerical models of solar-flare-like phenomena in space and laboratory plasmas.
\end{abstract}

\keywords{MHD---Sun: corona---Sun: magnetic fields} 

\section{INTRODUCTION}
	\label{s:intro}

In spite of the progress reached so far in understanding of the key role of magnetic reconnection in many phenomena of space and laboratory plasmas, its quantitative description in realistic three-dimensional (3D) configurations still remains a challenge \citep{Priest2000, Schindler2006}.
This concerns, in particular, the global aspects of reconnection such as an identification of reconnected magnetic fluxes and related structural changes in solar coronal configurations.
There are basically two fundamental reasons why this important problem is so complicated in the 3D case.

First, by analogy with the two-dimensional (2D) case, one could assume that partitioning the configurations into distinct interacting magnetic fluxes by separatrix surfaces (SSs) and estimating their time variation might help to resolve this problem.
And, indeed, this idea works very well whenever such SSs are due to the presence of magnetic null points (NPs) in the solar corona---the NP SSs are boundless surfaces, and so they provide the required partition.
However, the situation is different if the SSs are due to the presence of so-called bald patches (BPs), which are the segments of the photospheric polarity inversion line where coronal field lines touch the photosphere \citep{Seehafer1986a, Titov1993, Bungey1996, Delannee1999, Aulanier2002a}.
The BP SSs are usually not boundless and, therefore, not able to provide the required partition of the respective configurations.
It is also important that 
they are not exceptional structures, but are highly likely to be present in practice in the complex configurations that exist
in the solar atmosphere \citep{Titov1993, Bungey1996, Titov1999, Delannee1999, Aulanier2002a, Aulanier2002b, Pariat2004}.
Moreover, the set of 3D structural features where reconnection may occur is even wider than mentioned above, because it includes also quasi-separatrix layers (QSLs) \citep{Demoulin1996, Priest1995}.
Similar to BP SSs, these QSLs also do not provide an exact partition of configurations into distinct magnetic fluxes.

The second fundamental difficulty in quantitative description of 3D reconnection is related to magnetic diffusion and the breakdown of frozen-flux behavior.
Being an inherent part of the reconnection itself, magnetic diffusion in 3D can occur throughout the plasma volume in evolving configurations.
This is of particular importance for numerical MHD simulations, which may presently be performed only with a relatively high numerical or physical diffusion compared to real solar conditions.
Deviations from the frozen-in law condition are accumulated in this case throughout the volume to form a background against which it becomes difficult to distinguish between magnetic diffusion and reconnection. 

Nevertheless, significant progress in understanding 3D reconnection has been achieved in the framework of general magnetic reconnection (GMR) theory \citep{Schindler1988, Hesse1988, Hesse1991, Hesse1993}, which is focused on the analysis of magnetic field evolution constrained only by Maxwell's equations and Ohm's law.
On the basis of this theory, \citet{Hesse2005} have recently shown that, regardless of the presence of certain structural magnetic features, the time change of the reconnected magnetic flux is always directly related to the maximum of the field-line integral of parallel electric field.
For a plasma-magnetic configuration contained in a closed volume, this means that the change of the magnetic connection between the boundary plasma elements is fully controlled by the voltage drop along these lines.
It is clear, however, that such a change of the magnetic connection can always be found without calculating the electric field if the magnetic field in the volume and the plasma flows on the boundary are known at every moment.
As will be shown, the theory based on this premise complements the GMR theory by providing new intriguing opportunities for the analysis of magnetic reconnection.
The aim of this paper is to develop a general mathematical formalism that allows one to describe the change of magnetic connectivity and identify the reconnecting magnetic fluxes in any evolving configuration.
The new theory has been outlined earlier by \citet{Titov2007d}.
This paper, however, provides its full exposition.

The new formalism can also be viewed as a generalization of the theory of coronal magnetic connectivity, whose spatial variation is measured with the help of the so-called squashing factor $Q$.
This quantity was introduced first for closed magnetic configurations with a plane photospheric boundary \citep{Titov1999a, Titov2002a, Titov2002} and then extended to other configurations with arbitrary shapes of the boundaries \citep{Titov2007a}.
The $Q$ factor allows one to identify in a given configuration at a fixed time all the structural magnetic features, such as SSs and QSLs as well as their hybrids.
Here we generalize this quantity in order to characterize the evolution of magnetic connectivity by applying the $Q$ factor to so-called slip mapping---the mapping that describes the slippage between boundary plasma elements and footpoints of magnetic field lines.

It was previously noticed by \citet{Forbes1984a} for 2D magnetic configurations that the speed with which the photosphere is swept by a magnetic separatrix line allows one to estimate the electric field at the reconnection site.
\citet{Forbes2000} have also proposed strong arguments that this principle can be extended to 3D as well, excluding the above-mentioned configurations with QSLs, if one focuses only on the amount of magnetic flux reconnected up to a particular instant of time.
Our paper provides exactly this type of generalization of the principle for 3D configurations with QSLs included.

Section \ref{s:evol} describes the general theory of evolution of magnetic connectivity, which is illustrated in section \ref{s:exmpl1} for the example of magnetic evolution proposed earlier by \citet{Hesse2005}.
{{\hl
Section \ref{s:exmpl2} presents another example of magnetic evolution in a current layer configuration to show the difference between reconnection and magnetic diffusion.
}}
In section \ref{s:min}, we discuss the general properties of magnetic minimum points, which happen to be closely related to the considered theory.
The obtained results are summarized in section \ref{s:sum}.

%
%
\section{EVOLUTION OF MAGNETIC CONNECTIVITY
	\label{s:evol}}

\begin{figure}[htbp]
\epsscale{0.6}
\plotone{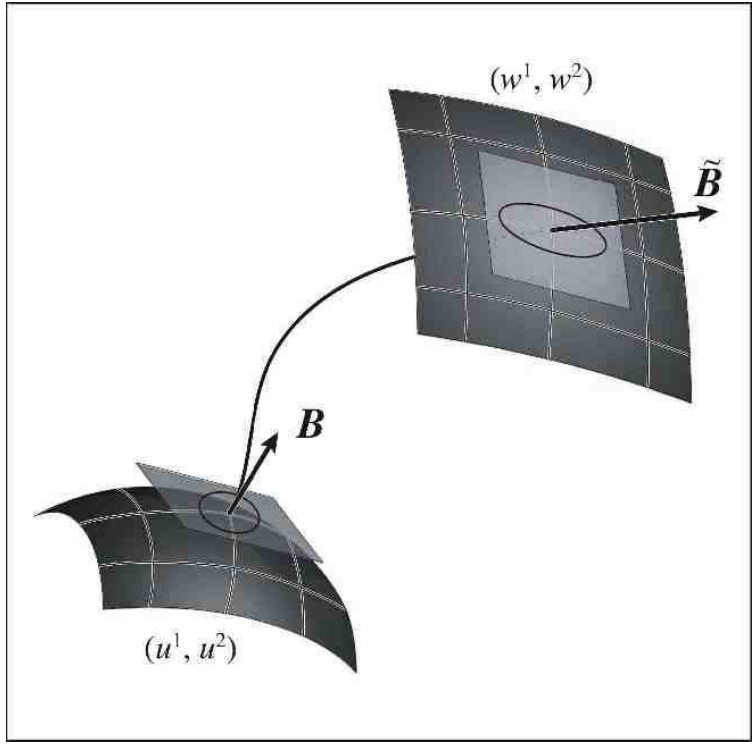}\\
\caption{Circle is mapped into an ellipse by a linearized field-line mapping acting between the tangent planes of the launch and target boundaries, where two different curvilinear coordinates $(u^{1},u^{2})$ and $(w^{1},w^{2})$, respectively, exist.  The aspect ratio of the ellipse, when it is large, coincides with a high value of the squashing factor~$Q$.
	\label{f:f1} }
\end{figure}

The basic concept that makes it possible to define the frozen-in law and its violation in magnetized-plasma flows is the connection of plasma elements by magnetic field lines \citep{Axford1984}.
When trying to characterize the magnetic structure and its evolution, we find it extremely useful to apply this concept to the plasma elements that are currently located at the boundary of a given configuration.
The magnetic field lines threading the plasma volume generally start and end up at some pieces of the boundary with the coordinate charts $(u^{1},u^{2})$ and $(w^{1},w^{2})$ (Figure \ref{f:f1}). 
In this way, they define a mapping $(u^{1},u^{2}) \rightarrow (w^{1},w^{2})$ determined by some vector function $(W^{1}(u^{1},u^{2}), W^{2}(u^{1},u^{2}))$ from one piece of the boundary to the other.
The local properties of this mapping are described by the Jacobian matrix
 \begin{eqnarray}
  D = \left[
         \frac{\partial W^{i}}{\partial u^{j}} 
      \right] .
  \label{D}
 \end{eqnarray}
For each field line, this matrix determines a linear mapping from the tangent plane at the launch footpoint to the tangent plane at the target footpoint, so that a circle in the first plane is mapped into an ellipse in the second plane (Figure \ref{f:f1}).
The aspect ratio $\lambda$ of such an ellipse defines the degree of local squashing  of elemental flux tubes, which means that any infinitesimal circle centered at a given launch point is mapped along the field lines into an infinitesimal ellipse with this aspect ratio at the target footpoint.

The value $\lambda$ is invariant with respect to the direction of the field-line mapping and so it can be used for characterizing the spatial variation of magnetic connectivity.
It turns out, however, that $Q\equiv\lambda + \lambda^{-1}$ provides a more compact expression in terms of coordinates and practically coincides with $\lambda$ in the most interesting cases where it is large.
So it is natural to use the value $Q$, called the squashing degree \citep{Titov2002} or squashing factor, for characterizing the local properties of magnetic connection between the boundary elements.  
The $Q$ factor can be expressed in terms of the norm $N$ \citep{ Priest1995, Demoulin1996} and Jacobian $\Delta$ of the field-line mapping as follows
\begin{eqnarray}
  Q & = & N^{2} /  |\Delta| .
  \label{Q}
\end{eqnarray}
It is convenient for further consideration to rewrite the covariant expressions for $N$ and $\Delta$ found by \citet{Titov2007a} for an arbitrary shape of the boundary in the following matrix form
\begin{eqnarray}
  N^{2} & = & \tr \left(D^{\mathrm T} G^{*} D\, G^{-1} \right) ,
	\label{N2}
\\
	|\Delta| & = & \left| \det \left(D^{\mathrm T} G^{*} D\, G^{-1} \right) \right|^{1/2} =
	|\det D| \left( \frac{g^{*}}{g} \right)^{1/2}
	       \equiv \left| \frac{B_{n}}{B^{*}_{n}} \right| ,
	\label{Dlt}
\end{eqnarray}
where $B_{n}$ and $B^{*}_{n}$ are the normal components of the magnetic field at the conjugate launch and target footpoints, respectively, and
\begin{eqnarray}
  G = \left[\frac{\partial {\bm R}}{\partial u^{i}}
            {\bm\cdot}
            \frac{\partial {\bm R}}{\partial u^{j}}
  \right]
	\label{G}
\end{eqnarray}
is the matrix of the covariant metric at the launch footpoints with $\det G$ denoted by $g$.
The radius-vector ${\bm R}(u^{1},u^{2})$ defining the shape of the boundary in a chosen coordinate chart $(u^{1},u^{2})$ is assumed here to be known.
The latter equally refers to the boundary at the target footpoints with radius-vector ${\bm R}(w^{1},w^{2})$, where the matrix of the covariant metric is
\begin{eqnarray}
  G^{*} = \left[
            \frac{\partial {\bm R}}{\partial w^{i}}
            {\bm\cdot}
            \frac{\partial {\bm R}}{\partial w^{j}}
  \right]^{*} .
	\label{G*}
\end{eqnarray}
The asterisk symbol (superscript) indicates here that the values of $G$ refer to the conjugate target footpoints such that $(w^{1},w^{2}) = (W^{1}(u^{1},u^{2}), W^{2}(u^{1},u^{2}))$.
In other words, the asterisk symbol pulls back the values of $G$ from the target to the launch footpoints at which the $Q$ factor is evaluated.

If the magnetic field ${\bm B}({\bm r})$ is defined analytically or on some numerical grid, the Jacobian matrix (\ref{D}) can be straightforwardly computed from the coordinates of the target footpoints that are obtained by integrating the field-line equation
\begin{eqnarray}
  \frac{{\mathrm d}{\bm r}}{{\mathrm d}\tau}  =  {\bm B}  
	\label{FLE}
\end{eqnarray}
with the initial condition at launch footpoints
\begin{eqnarray}
  \left.{\bm r}\right|_{\tau=0} =  {\bm R}(u^{1},u^{2}) 
	\label{ICFLE} 
\end{eqnarray}
and the stopping condition at target footpoints
\begin{eqnarray}
	\left.{\bm r}\right|_{\tau=\tau_{\mathrm s}} =  {\bm R}(w^{1},w^{2}) , 
	\label{SCFLE}
\end{eqnarray}
which are satisfied by some parameter values $\tau_{\mathrm s}$ to be found.
Tracing of the field lines should be done with a sufficiently high accuracy to make the subsequent calculation of the coordinate derivatives accurate enough.
Although such computations are relatively time consuming, they may be fulfilled with a very good resolution even for the entire solar corona without using supercomputers \citep{Titov2008a}.
The latter are required only for numerical modeling of the coronal magnetic field itself.
For example, such a computation for the given magnetic field data on a Mac Pro with one processor Intel Xeon (2GHz) requires only 23 hr.
The resulting $Q$ factor is obtained on a uniform grid with 3600 and 7200 mesh points in latitude and longitude, respectively, for a magnetic field defined in spherical coordinates $(r, \theta, \phi)$ on a nonuniform grid with 174 x 118 x 174 mesh points.

As mentioned in section \ref{s:intro}, the $Q$ factor allows one to identify in a given configuration all topological and geometrical magnetic features at a fixed time instant.
Formally, the SSs are distinguished from the QSLs by infinite rather than large finite values of $Q$.
However, from a numerical point of view, these features are similar, since they are both identified via large finite values of $Q$.
The only difference is that the maxima of $Q$ increase at SSs limitlessly with a decreasing size of a numerical grid used for computing $Q$, while at QSLs such maxima converge to certain large values.
Such a subtle difference is usually not important for applications and so, for brevity, we will use hereafter the term QSLs in a broad sense to include also SSs, unless it is stated otherwise.

%
%
\subsection{Ideal evolution of the magnetic field \label{s:ideal}}

\begin{figure}[htbp]
\epsscale{0.6}
\plotone{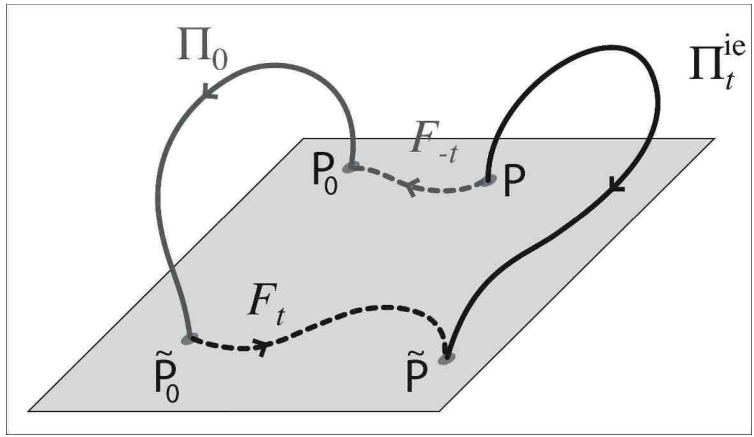}
\caption{Field-line mapping $\Pi^{\mathrm{ie}}_{t}$  is expressed
at any time $t$ in terms of the initial field-line mapping $\Pi_{0}$ and the tangential forward and backward boundary flows $F_{t}$ and $F_{-t}$, respectively [see equation (\ref{Pit})]. This mapping occurs whenever the frozen-in law is valid for the evolution of the configuration. The field lines and trajectories of the footpoints are shown by solid and dashed lines, respectively.
	\label{f:f2} }
\end{figure}

If the evolution of magnetic configurations is subject to the frozen-in law, the initial connection of plasma elements by magnetic field lines remains unchanged.
The analysis of magnetic connectivity then is significantly simplified, because the field-line mapping $\Pi^{\mathrm{ie}}_{t}$ at any time $t$ can be composed in this case as follows: 
%
\begin{eqnarray}
  \Pi^{\mathrm{ie}}_{t} = F_{t}\circ \Pi_{0} \circ F_{-t} ,
	\label{Pit}
\end{eqnarray}
where the superscript ``ie" stands for the ``ideal evolution", $\Pi_{0}$ is the initial field-line mapping, while $F_{t}$ and $F_{-t}$ are tangential boundary flows, forward and backward in time, respectively (see Figure \ref{f:f2}).
This means that for any footpoint P, its conjugate footpoint $\tilde{\mathrm P}\equiv\Pi^{\mathrm{ie}}_{t}({\mathrm P})$ can generally be found by following the lower path of the diagram
%
\begin{eqnarray}
\begin{array}{lll}
{\mathrm P}     & \stackrel{\scriptscriptstyle\Pi^{\mathrm{ie}}_{t}}{\longrightarrow} & \tilde{\mathrm P}\\
\downarrow {\scriptscriptstyle F_{-t}} &\hspace{2.0em} & \uparrow {\scriptscriptstyle F_{t}} \\
{\mathrm P}_{0} & \stackrel{\scriptscriptstyle\Pi_{0}}{\longrightarrow} & \tilde{\mathrm P}_{0}
\end{array}
\end{eqnarray}
In other words, instead of tracing the field line that connects P and $\tilde{\mathrm P}$ at time $t$ (see Figure \ref{f:f2}), one can first trace the trajectory of P backward in time to find its initial prototype ${\mathrm P}_{0}$, then tracing from it the field line of the initial configuration find its conjugate prototype $\tilde{\mathrm P}_{0}$, and finally tracing the trajectory from $\tilde{\mathrm P}_{0}$ forward in time find $\tilde{\mathrm P}$.
Such a three-step calculation of the field-line mapping allows one to avoid the use of magnetic field data except for the initial data.
Thus, whenever the tangential boundary flows and the initial magnetic field are known, the squashing factor may be calculated for any time moment without computing the ideal MHD evolution of configuration itself.

This is very important if one takes into account the results of numerical MHD simulations \citep{Longcope1994a, Aulanier2005, Titov2008a} that demonstrate a good correlation between regions with a highly distorted field-line mapping and strong current concentrations.
In the light of this, we conjecture that equation (\ref{Pit}) must provide a powerful tool for predicting favorable sites of current-layer formation in magnetic configurations whose evolution is subject to the frozen-in law.
Its first successful application to a particular configuration with two twisted magnetic flux spots \citep{Titov2008a} strongly supports this conjecture.

In coordinate notations, the field-line mapping defined by equation (\ref{Pit}) is described as
%
\begin{eqnarray}
 \Pi^{\mathrm{ie}}_{t}: (u^{1},u^{2})
    \stackrel{\overbrace{\scriptscriptstyle (U^{1}_{0}, U^{2}_{0})}^{F_{-t}} }{\longrightarrow} 
                                 (u^{1}_{0}, u^{2}_{0})
    \stackrel{\overbrace{\scriptscriptstyle(W^{1}_{0},  W^{2}_{0})}^{\Pi_{0}} }{\longrightarrow} 
                                (w^{1}_{0}, w^{2}_{0}) 
    \stackrel{\overbrace{\scriptscriptstyle(W^{1}    ,  W^{2}    )}^{F_{t}} }{\longrightarrow}
                                (w^{1}    , w^{2}    ) ,
	\label{Ptc}
\end{eqnarray}
where the coordinate functions and their values are denoted by the same letters but different cases---the upper-case letters represent the functions, while the low-case letters represent their values; the functions and coordinates referring to the initial time moment are marked with the subscript 0.
Thus, for example, the flow $F_{-t}$ maps $(u^{1}, u^{2})$ to ($u^{1}_{0} = U^{1}_{0}(u^{1}, u^{2})$, $u^{2}_{0} = U^{2}_{0}(u^{1}, u^{2})$) and so on.
The second step in this composition requires solving of the problem described by equations (\ref{FLE})--(\ref{SCFLE}), with the proviso that this problem is set for the initial configuration, whose boundary coordinates are $(u^{1}_{0},u^{2}_{0})$ and $(w^{1}_{0},w^{2}_{0})$. 
The first and third steps of composition (\ref{Ptc}) require tracing of the footpoint trajectories, which are determined from the tangential components of the boundary velocity field ${\bm v}_{\mathrm b}$, assumed to be known.
Specifically, one needs to solve the following two initial value problems.
For the first step, the system
\begin{eqnarray}
  \frac{{\mathrm d}U^{i}_{0}}{{\mathrm d}\tilde{t}} & = & g^{is} 
     \left(\frac{\partial {\bm R}}{\partial u^{s}} 
       \bm{\cdot} {\bm v}_{\mathrm b} \right), \quad i=1,2; \: s=1,2; 
	\label{U0eq}  \\
	  \left. (U^{1}_{0},U^{2}_{0})\right|_{\tilde{t}=t} &=& (u^{1},u^{2}),
	\label{U0IC}
\end{eqnarray}
should be integrated {\em backward} in time from $t$ to 0,
while, for the third step, the system
\begin{eqnarray}
  \frac{{\mathrm d}W^{i}}{{\mathrm d}\tilde{t}} & = & g^{is} 
     \left(\frac{\partial {\bm R}}{\partial w^{s}} 
       \bm{\cdot} {\bm v}_{\mathrm b} \right), \quad i=1,2; \: s=1,2; 
	\label{Weq}  \\
	  \left. (W^{1},W^{2})\right|_{\tilde{t}=0} &=& (w^{1}_{0},w^{2}_{0})^{\ad},
	\label{WIC}
\end{eqnarray}
should be integrated {\em forward} in time from $0$ to $t$.
In both these systems, summation over the repeating index $s$ is assumed;
the contravariant metric $\left[g^{is}\right] \equiv G^{-1}$ refers to points of the launch and target boundaries for equations (\ref{U0eq}) and (\ref{Weq}), respectively.
The double-asterisk symbol (in the superscript) in equation (\ref{WIC}) indicates that the point $(w^{1}_{0},w^{2}_{0})$ is the image of $(u^{1},u^{2})$ that is obtained at the first two steps of composition (\ref{Ptc}).
In other words, the double asterisks provide a double pull back of this point to the launch boundary, where $Q$ is evaluated.
This convention will be used hereafter for other values as well, with the extension to a triple pull back denoted by triple asterisks, respectively.

It is clear that the first and third steps in the procedure described above are no more difficult than the second step.
In principle, making consecutively these three steps at least three times in the small neighborhood of a given launch point, one can evaluate numerically the Jacobian matrix (\ref{D}) and the respective value of $Q$ [see eqs. (\ref{Q})--(\ref{G*})].
In practice, however, the divergence of the field lines at QSLs of the initial configuration can be so high that it may significantly affect the accuracy of such an evaluation.
A better control of the accuracy in this procedure is obtained by using the fact that the Jacobian matrix $D_{\mathrm{ie}}$ of the composite``ideal" mapping (\ref{Pit}) is the product of the Jacobian matrices of the individual mappings, namely
\begin{eqnarray}
  D_{\mathrm{ie}} = 
  \underbrace{
    \left[ \frac{\partial W^{i}}{\partial w^{p}_{0}} \right]^{\ad} }_{M^{\ad}}
  \underbrace{
    \left[ \frac{\partial W^{p}_{0}}{\partial u^{s}_{0}} \right]^{*} }_{D^{*}_{0}}
  \underbrace{
    \left[ \frac{\partial U^{s}_{0}}{\partial u^{j}}\right] }_{M^{-1}} . 
	\label{Dc}
\end{eqnarray}
Thus, the calculation of $D$ is reduced to the calculations of these three matrices at the points that ``lie on the path" of the composite mapping.
As stated above, each of these calculations is similar to one another, and none of them requires knowledge of the magnetic field, except at the initial moment.

It is clear from this consideration that the respective squashing factor $Q_{\mathrm{ie}}$ can be obtained now by modifying equations (\ref{Q})--(\ref{Dlt}) as follows:
\begin{eqnarray*}
  Q_{\mathrm{ie}}  = \frac{ 
   \tr \left(D^{\mathrm T}_{\mathrm{ie}}\, G^{\aaa}\, 
             D_{\mathrm{ie}}            \, G^{-1} 
              \right)
        } 
        {\left|\det \left(D^{\mathrm T}_{\mathrm{ie}}\, G^{\aaa}\, 
                          D_{\mathrm{ie}}            \, G^{-1} 
              \right)\right|^{1/2}
        }.
\end{eqnarray*}
The determinant of the product of the individual matrices  in this expression can be rewritten as a product of the corresponding individual determinants.
Using also conservation of the magnetic flux in the initial configuration, we finally obtain
\begin{eqnarray}
Q_{\mathrm{ie}}  = 
    \left(\frac{g^{\ad}_{0} g}{g^{*}_{0}\, g^{\aaa}}\right)^{1/2}
    \left|\frac{B^{\ad}_{n0}}{B^{*}_{n0}}\right|
   \frac{ 
   \tr \left(D^{\mathrm T}_{\mathrm{ie}}\, G^{\aaa}\, 
             D_{\mathrm{ie}}            \, G^{-1} 
              \right)
        } 
        {\det(M^{\ad} \, M^{-1})
        }.
	\label{Qie}
\end{eqnarray}
We used here the fact that $\det D^{*}_{0} = B^{*}_{n0}\sqrt{g^{*}_{0}}/(B^{\ad}_{n0} \sqrt{g^{\ad}_{0}})$, where $B^{*}_{n0}$ and $B^{\ad}_{n0}$ are the normal components of the initial field at the corresponding conjugate footpoints, with $g^{*}_{0} \equiv \det G^{*}_{0}$ and $g^{\ad}_{0} \equiv \det G^{\ad}_{0} $ at such points.

%
%
\subsection{Nonideal evolution of the magnetic field \label{s:nonideal}}

\begin{figure}[htbp]
\epsscale{0.6}
\plotone{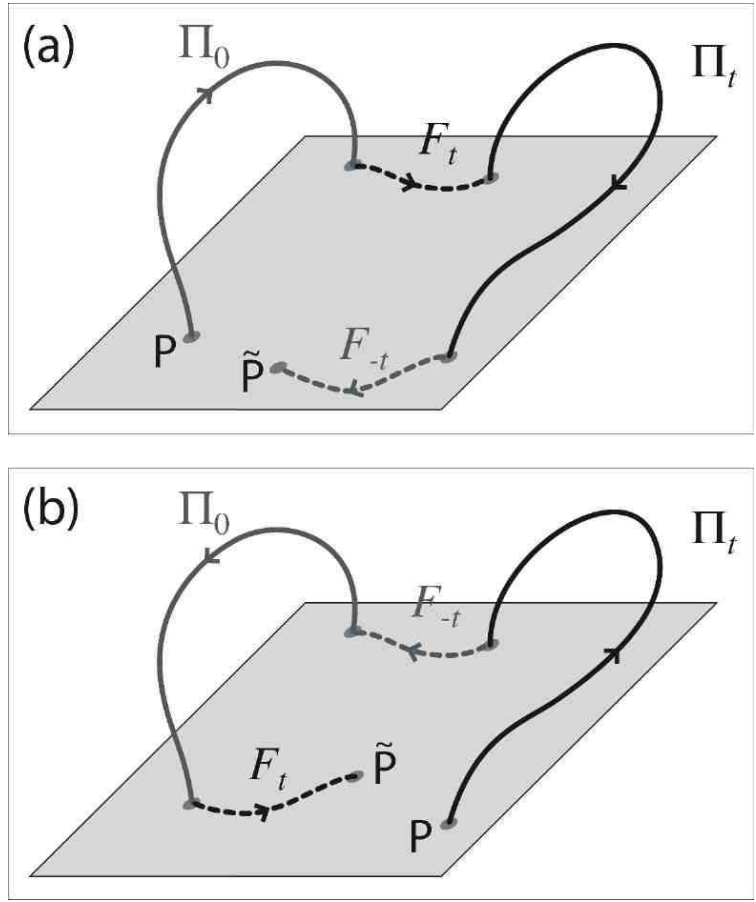}
\caption{Slip-forth (a) and slip-back (b) mappings [see equations (\ref{sfm}) and (\ref{sbm}), respectively] act on the footpoints of the field lines in such a way that each footpoint P and its image $\tilde{\mathrm P}$ are generally different due to a breakdown of the frozen-in condition.
	\label{f:f3} }
\end{figure}

The breaking of the frozen-in law causes a change of the initial magnetic connection of the boundary plasma elements.
In other words, such elements experience, in this case, a slippage relative to their ideal MHD mapping.
This intuitive notion can be formalized by introducing two types of mapping, which relate two time moments 0 and $t$ in the evolution of a given magnetic configuration.
The first of them, denoted by $S_{t}$ and expressed as
%
\begin{eqnarray}
  S_{t} = F_{-t}\circ \Pi_{t} \circ F_{t}\circ\Pi_{0} ,
	\label{sfm}
\end{eqnarray}
maps the boundary onto itself at the initial time $0$.
We will call this the {\it slip-forth mapping} because the slippage of footpoints occurs {\it forward} in time.
The second mapping describes a corresponding slippage {\it backward} in time and it is expressed as
%
\begin{eqnarray}
  S_{-t} = F_{t}\circ \Pi_{0} \circ F_{-t}\circ\Pi_{t}
  \label{sbm}
\end{eqnarray}
and called the {\it slip-back mapping}.
This expression also maps the boundary onto itself but at the final time $t$, so that the slip-back mapping is an antipode to the slip-forth mapping.
A pointwise definition of these mappings is depicted in Figure \ref{f:f3} for the case of a closed magnetic configuration.
The picture can be easily modified, however, to represent more general configurations comprising both closed and open field lines by simply enclosing the coronal volume under study with the help of an additional upper boundary surface.

Comparing equations (\ref{sbm}) and (\ref{Pit}), we obtain
%
\begin{eqnarray}
  S_{-t} = \Pi^{\mathrm{ie}}_{t} \circ \Pi_{t},
	\label{sbm_sh}
\end{eqnarray}
which means that the slip-back mapping is just a composite of the ``real" and ``ideal" field-line mappings at time $t$.
Similarly, equation (\ref{sfm}) can be rewritten as
\begin{eqnarray}
  S_{t} & = & \Pi^{\mathrm{ie}}_{0} \circ \Pi_{0},
	\label{sfm_sh}
\end{eqnarray}
where
\begin{eqnarray}
	\Pi^{\mathrm{ie}}_{0} & = & F_{-t}\circ \Pi_{t} \circ F_{t}
	\label{Pi0}
\end{eqnarray}
is that initial field-line mapping whose ideal evolution driven by the flow $F_{t}$ would produce at time $t$ the final field-line mapping $\Pi_{t}$.
Thus, both slip mappings identify the differences in the field-line connectivities of the ``real" and ``ideal" configurations.
The slip-forth mapping $S_{t}$, however, does this for the initial observer, while the slip-back mapping $S_{-t}$ does the same for the final observer.

This consideration also shows that the footpoint slippage that  will occur or has occurred within time $t$ is fully determined by the magnetic fields at the initial and final moments and by the respective displacements of the plasma elements due to tangential boundary flows.
Such a slippage is caused by both magnetic reconnection and resistive plasma diffusion, whose contributions may generally be comparable in value.
Yet it is clear that magnetic reconnection is more localized in space than resistive diffusion \citep{Schindler1988}.
One can expect therefore that these two processes are discriminated not by the magnitude of the slippage itself but rather by its spatial gradient.
Note also that the introduced slip mappings are very similar to the field-line mapping in the sense that all of them are just two-dimensional mappings of the boundary on itself.
Thinking by analogy, we can assume then that the squashing factor is a major characteristic not only for a field-line mapping but also for slip mappings.
This assumption is fully confirmed by our further analysis, which shows that in the case of slip mappings the squashing factor makes it possible to identify the reconnecting magnetic flux tubes in any evolving configuration, even if its evolution involves a substantial resistive diffusion.
{{\hl
In particular, section \ref{s:exmpl2} presents an example of magnetic evolution in a current layer with a small nonideal region inside.
It shows that the transition from magnetic diffusion to reconnection is not abrupt but continuous.
The reconnecting flux tubes here pass through, and in the vicinity of, the nonideal region and they are distinguished, as expected, by high values of the squashing factor that are related to a locally enhanced current density.   
}}

To derive the required expressions, let us first represent the slip mappings in coordinate notations.
For the slip-forth mapping defined by equation (\ref{sfm}), we have
\begin{eqnarray}
 S_{t}: (u^{1}_{0},u^{2}_{0})  
    \stackrel{\overbrace{\scriptscriptstyle (W^{1}_{0}, W^{2}_{0})}^{\Pi_{0}}} {\longrightarrow}                           (w^{1}_{0}, w^{2}_{0})
    \stackrel{\overbrace{\scriptscriptstyle (W^{1}    , W^{2}    )}^{  F_{t}}} {\longrightarrow}                           (w^{1}    , w^{2}    )
    \stackrel{\overbrace{\scriptscriptstyle (U^{1}    , U^{2}    )}^{\Pi_{t}}} {\longrightarrow}                           (u^{1}    , u^{2}    )
    \stackrel{\overbrace{\scriptscriptstyle (U^{1}_{0}, U^{2}_{0})}^{ F_{-t}}} {\longrightarrow}                           (u^{1}_{0}, u^{2}_{0}) ,
	\label{sfm_c}
\end{eqnarray}
so that its Jacobian matrix is determined by the following product of individual Jacobian matrices:
%
\begin{eqnarray}
  D_{\mathrm{sf}} \equiv 
    \left[\frac{\partial U^{i}_{0}}{\partial u^{j}_{0}}\right] =
  \underbrace{
    \left[\frac{\partial U^{i}_{0}}{\partial u^{p}    }\right]^{\aaa}}
    _{M^{-1\aaas}}
  \underbrace{
    \left[\frac{\partial U^{p}    }{\partial w^{q}    }\right]^{\ad}}
    _{D^{\ad}}
  \underbrace{ 
    \left[\frac{\partial W^{q}    }{\partial w^{s}_{0}}\right]^{*  }}
    _{M^{*}}
  \underbrace{
    \left[\frac{\partial W^{s}_{0}}{\partial u^{j}_{0}}\right]^{   }}
    _{D_{0}}
	\label{Dsf}
\end{eqnarray}
This composition implies that the respective squashing factor $Q_{\mathrm{sf}}$ can be derived from suitably modified equations (\ref{Q})--(\ref{Dlt}) as follows:
\begin{eqnarray*}
  Q_{\mathrm{sf}} = 
    \frac{\tr \left( D^{\mathrm T}_{\mathrm{sf}} \, G^{\aaaa}_{0} \, 
                     D_{\mathrm{sf}}\, G^{-1}_{0} \right)}
    {\left|
          \det \left( D^{\mathrm T}_{\mathrm{sf}} \, G^{\aaaa}_{0} \, D_{\mathrm{sf}}\, G^{-1}_{0} \right) \right|^{1/2}} .
\end{eqnarray*}
Rewriting the denominator in this expression as a product of individual determinants of the entering matrices and using conservation of a magnetic flux in the initial and final configurations, we finally obtain
\begin{eqnarray}
  Q_{\mathrm{sf}} =
    \left( \frac{g^{\aaa} g^{*}_{0}}{g^{\ad} g^{\aaaa}_{0}} \right)^{1/2}
    \left| \frac{B^{\aaa}_{n} B^{*}_{n0}}{B^{\ad}_{n} B_{n0}} \right|
    \frac{\tr \left( D^{\mathrm T}_{\mathrm{sf}} \, G^{\aaaa}_{0} \, D_{\mathrm{sf}} \, G^{-1}_{0} \right)}{ \det\left( M^{-1\aaa} \, M^{*} \right)} .
	\label{Qsf}
\end{eqnarray}

Similarly, the slip-back mapping defined by equation (\ref{sbm}) is represented
in coordinates by
\begin{eqnarray}
 S_{-t}: (u^{1},u^{2})  
    \stackrel{\overbrace{\scriptscriptstyle (W^{1}    , W^{2}    )}^{\Pi_{t}}} {\longrightarrow}                           (w^{1}    , w^{2}    )
    \stackrel{\overbrace{\scriptscriptstyle (W^{1}_{0}, W^{2}_{0})}^{F_{-t} }} {\longrightarrow}                           (w^{1}_{0}, w^{2}_{0})
    \stackrel{\overbrace{\scriptscriptstyle (U^{1}_{0}, U^{2}_{0})}^{\Pi_{0}}} {\longrightarrow}                           (u^{1}_{0}, u^{2}_{0})
    \stackrel{\overbrace{\scriptscriptstyle (U^{1}    , U^{2}    )}^{F_{t}}} {\longrightarrow}                           (u^{1}    ,  u^{2}    ) ,
	\label{sbm_c}
\end{eqnarray}
so that its Jacobian matrix is determined by the following product of individual Jacobian matrices:
%
\begin{eqnarray}
  D_{\mathrm{sb}} \equiv
    \left[\frac{\partial U^{i}    }{\partial u^{j}    }\right] =
  \underbrace{
    \left[\frac{\partial U^{i}    }{\partial u^{p}_{0}}\right]^{\aaa}}
  _{M^{\aaas}}
  \underbrace{
    \left[\frac{\partial U^{p}_{0}}{\partial w^{q}_{0}}\right]^{\ad}}
  _{D_{0}^{\ad}}
  \underbrace{
    \left[\frac{\partial W^{q}_{0}}{\partial w^{s}    }\right]^{*  }}
  _{M^{-1*}}
  \underbrace{
    \left[\frac{\partial W^{s}    }{\partial u^{j}    }\right]      }
  _{D} .
	\label{Dsb}
\end{eqnarray}
This composition implies that the respective squashing factor $Q_{\mathrm{sb}}$ can be obtained from suitably modified equations (\ref{Q})--(\ref{Dlt}) as
\begin{eqnarray*}
  Q_{\mathrm{sb}} = 
    \frac{\tr \left( D^{\mathrm T}_{\mathrm{sb}} \, G^{\aaaa} \, 
                     D_{\mathrm{sb}}\, G^{-1} \right)}
    {\left|
          \det \left( D^{\mathrm T}_{\mathrm{sb}} \, G^{\aaaa} \, D_{\mathrm{sb}}\, G^{-1} \right) \right|^{1/2}} .
\end{eqnarray*}
Rewriting the denominator in this expression as a product of individual determinants of the entering matrices and using conservation of a magnetic flux in the initial and final configurations, we finally obtain
\begin{eqnarray}
  Q_{\mathrm{sb}} =
    \left( \frac{g^{\aaa}_{0} g^{*}}{g^{\ad}_{0} g^{\aaaa}} \right)^{1/2}
    \left| \frac{B^{\aaa}_{n0} B^{*}_{n}}{B^{\ad}_{n0} B_{n}} \right|
    \frac{\tr \left( D^{\mathrm T}_{\mathrm{sb}} \, G^{\aaaa} \, D_{\mathrm{sb}}\, G^{-1} \right)}{ \det\left( M^{\aaa}\, M^{-1*} \right)}  .
	\label{Qsb}
\end{eqnarray}
%

By construction, the $Q_{\mathrm{sf}}$ factor is defined on the boundary of the {\it initial} magnetic configuration by its {\it subsequent} evolution within time $t$,
while the $Q_{\mathrm{sb}}$ factor is defined on the boundary of the {\it final} magnetic configuration by its {\it preceding} evolution within time $t$.
Both slip-squashing factors depend only on the two Jacobian matrices of the field-line mapping, $D_0$ and $D$, corresponding, respectively, to the initial and final moments.

To better understand the meaning of these characteristics, consider a situation where the variation of the footpoint positions due to nonideal processes in the volume is much larger than that due to the surface flows at the boundary.
From the physical point of view, this is actually the most interesting situation, because it corresponds to the case where magnetic reconnection going on in the volume is so quick on the time scale under study that the boundary motion of plasma can be neglected.
The whole paths of the slip mappings are formed in this case only by field lines of the initial and final configurations, while $F_{t}$ and $F_{-t}$ in (\ref{sfm})--(\ref{sbm}) are approximately identity mappings characterized by unit Jacobian matrices.
Taking into account these observations, one can approximate equations (\ref{Qsf}) and (\ref{Qsb}) by
\begin{eqnarray}
  Q_{\mathrm{sf}} &\approx&   
    \left( \frac{g^{\ad} g^{*}_{0}}{g^{*} g^{\ad}_{0}} \right)^{1/2}
    \left| \frac{B^{\ad}_{n} B^{*}_{n0}}{B^{*}_{n} B_{n0}} \right|
    \tr \left( D^{\mathrm T}_{0} D^{\mathrm T*} \, G^{\ad}_{0} \, D^{*} D_{0} \, G^{-1}_{0} \right) ,
     \label{Qsfa}\\
  Q_{\mathrm{sb}} &\approx& 
    \left( \frac{g^{\ad}_{0} g^{*}}{g^{*}_{0} g^{\ad}} \right)^{1/2}
    \left| \frac{B^{\ad}_{n0} B^{*}_{n}}{B^{*}_{n0} B_{n}} \right|
    \tr \left(D^{\mathrm T} D^{\mathrm T *}_{0} \, G^{\ad} \, D^{*}_{0} D\, G^{-1} \right) ,  
	\label{Qsba}	
\end{eqnarray} 
which show that, in general, if some of the matrix elements in $D_0$ or $D$ are large in absolute value,  $Q_{\mathrm{sf}}$ or $Q_{\mathrm{sb}}$ should be large as well.
Such matrices, however, are related to the $Q$ factor of the initial or final magnetic field in a similar way.
Thus, $Q_{\mathrm{sf}}$ or $Q_{\mathrm{sb}}$ should be large at the mapping paths [see eqs. (\ref{sfm_c}) and (\ref{sbm_c})] that include the field lines belonging to the initial or final QSLs of the evolving configuration.

By analogy with the QSLs, one can expect that the flux tubes assembled from such field lines have a layer-like structure.
We will call these flux tubes {\it reconnection fronts} (RFs).
Their determination is of great importance for analyzing magnetic reconnection in complicated 3D configurations.
Such configurations, however, require special consideration, and so the discussion of them is postponed to section \ref{s:rec_ft}.
We note here only that both slip mappings have two RFs, one of which, called {\it the instantaneous RF}, corresponds to the present moment (0 or $t$, respectively, for the slip-forth or slip-back mappings).
Equations (\ref{Qsfa}) and (\ref{Qsba}) suggest that such a reconnection front should approximately coincide with the set of QSLs existing at a present moment (see an example in section \ref{s:exmpl1}).
However, the situation is different for the second RF, which is called {\it the future RF} or {\it past RF}, respectively, for the slip-forth or slip-back mapping.
The future RF consists of those initial flux tubes whose footprints will turn after reconnection into the footprints of the final QSLs.
The past RF consists of those final flux tubes whose footprints are formed by reconnection from the footprints of the initial QSLs. 
These definitions of RFs are slightly modified if the advection of footpoints is essential in the evolution under study.

%
%
\subsection{To-be-reconnected and reconnected magnetic fluxes \label{s:rec_ft}}

It might seem surprising at first glance that our description of magnetic reconnection does not invoke an electric field at all, even though the component of the electric field parallel to a magnetic field is at the cornerstone of the GMR theory \citep{Schindler1988, Hesse1988}.
The physical ground for this omission of the electric field becomes clear if one recalls that its parallel component appears generally in very strong current layers (CLs), in particular, when the corresponding resistive voltage drop is not negligible.
However, QSLs are likely to be favorable sites for the formation of strong CLs under very general conditions \citep{Longcope1994a, Priest1995, Titov2003, Galsgaard2003, Aulanier2005}.
This makes it possible to put QSLs at the core of general reconnection theory by reformulating it in purely geometrical terms.
Such a reformulation has the advantage of being able to directly relate the reconnection process with an accompanying variation of the structure of evolving magnetic configurations.
{{\hl
More details illustrating these general statements are given in sections \ref{s:exmpl1} and \ref{s:exmpl2}.
}}

What is particularly important is that our approach enables us to identify reconnecting      flux tubes in a given magnetic evolution as well as to estimate the reconnected fluxes by tracing the footprints of the RFs.
To describe this more precisely, let us introduce first the lines going along the ridges of the  $Q$ distributions by denoting such lines as $\QSSfp$ and $\RFfp$, respectively, for the field-line and slip mappings.
QSS abbreviates here a quasi-separatrix surface, which is a magnetic surface passing through the ridge of the respective $Q$ distribution at the boundary.
QSSs play the role of the middle surfaces for QSLs [see some examples of QSSs in \citep{Titov2008a}], so that $\QSSfp$ lines serve as midlines for the footprints of QSLs.
Similarly, the $\RFfp$ lines serve as midlines for the footprints of RFs.

Consider first the simplest case, where the advection of the footpoints is negligible in comparison with their displacements due to reconnection, which means that both $F_{t}$ and $F_{-t}$ are approximately identity mappings in composition (\ref{sfm_c}).
Assume in addition that there are QSLs at both time moments and that their footprints are sufficiently separated in distance.
This implies that the squashing of the $S_{t}$ mapping on its composite path passing through a given footprint of the QSL is mainly provided by the field-line mapping corresponding to this QSL.
By noticing also that $\Pi^{-1}_{0} = \Pi_{0}$, we obtain then from relation (\ref{sfm_c}) that
%
\begin{eqnarray}
  \RFfp^{\pm}_{0} & \approx & \QSSfp^{\pm}_{0} , 
	  \label{sf_RFfp_0}  \\
	\RFfp^{\pm}_{t} & \approx & \Pi_{0}(\QSSfp^{\mp}_{t}) ,
	  \label{sf_RFfp_ta1}
\end{eqnarray}
where the subscripts $0$ or $t$ correspond, respectively, to instantaneous or future RFs, while the superscripts $+$ or $-$ denote the positive or negative magnetic polarity, to which the $\RFfp$- or $\QSSfp$ line belongs.

These equalities establish a direct relationship between RFs and QSLs and provide a simplified algorithm for determining RFs in the described situation.
As discussed in section \ref{s:intro}, QSSs and boundary surfaces do not necessarily bound closed volumes in the corona, and so $\RFfp_{0}$- and $\RFfp_{t}$ lines might not form closed contours at the boundary.
The latter may happen, however, if QSSs, for example, turn out to be NP SSs.
In this case, the closed contours, formed by $\RFfp_{0}$- and $\RFfp_{t}$ lines, outline in the initial configuration the footprints of the flux tubes that will reconnect within time $t$.
We will call them {\it to-be-reconnected} magnetic flux tubes.
In a more general situation with open $\RFfp$ lines, the footprints of to-be-reconnected flux tubes are the boundary areas that are swept by $\RFfp_{t}$ lines in the initial configuration with time running from $0$ to $t$.
The integral of the normal component of the initial magnetic field over such areas determines the magnetic fluxes that will be reconnected within time $t$.

Thus, the $\RFfp$ lines play a crucial role in determining the reconnecting magnetic fluxes---the more accurately we determine the geometry and kinematics of the $\RFfp$ lines, the better we can quantify the reconnecting flux tubes for a given magnetic evolution.
Therefore, it is important to understand also what happens if the reconnection is not fast enough to neglect the advection of footpoints. 
To understand this question, let us turn again to equation (\ref{sfm_c}) and its more compact form (\ref{sfm_sh})--(\ref{Pi0}), from where one can see that the location of $\RFfp_{0}$ lines should be given by equation (\ref{sf_RFfp_0}) if the paths of $S_{t}$ [see eq. (\ref{sfm_c})], on which the field-line mappings $\Pi_{0}$ or $\Pi^{\mathrm{ie}}_{0}$ are mostly distorted, are different.
The latter is probably valid if the time interval $t$ under study is not too small and so the footprints of QSLs of the $\Pi_{0}$ or $\Pi^{\mathrm{ie}}_{0}$ mappings do not overlap each other.

Yet the situation with the $\RFfp_{t}$ lines is more complicated, even if the latter is true.
The answer in this case depends on whether or not the boundary flows $F_{t}$ and $F_{-t}$
significantly interlock or unlock the field lines in the volume of the final magnetic configuration.
This, in turn, depends on whether or not these flows contain substantial shearing motions applied in transverse directions to the conjugate footprints of QSLs.
Because a twisting pair of such shearing motions is able to seriously modify the $\Pi^{\mathrm{ie}}_{0}$ mapping compared to the $\Pi_{t}$ mapping.
In fact, this kind of advection can even destroy the original QSL or, in contrast, it can create a new one, depending on the sign of the twist and the resulting values of shear \citep{Titov2003, Galsgaard2003, Titov2008a}.
If such an interlocking or unlocking effect of the boundary flows is small (but the footpoint advection is still not negligible), the $\RFfp_{t}$ lines can be approximated by 
%
\begin{eqnarray}
		\RFfp^{\pm}_{t} & \approx & \Pi_{0} \circ F_{-t}(\QSSfp^{\mp}_{t}) .
	  \label{sf_RFfp_ta2}
\end{eqnarray}
Otherwise, one needs to use a more accurate approximation, namely,
%
\begin{eqnarray}
		\RFfp^{\pm}_{t} & \approx & \Pi_{0} (\widetilde{\QSSfp}^{\mp}_{0}) ,
	  \label{sf_RFfp_ta3}
\end{eqnarray}
where $\widetilde{\QSSfp}^{\mp}_{0}$ denotes the $\QSSfp$ lines of the $\Pi^{\mathrm{ie}}_{0}$ mapping defined by equation (\ref{Pi0}).
Thus, the sequence of equations (\ref{sf_RFfp_ta1}), (\ref{sf_RFfp_ta2}) and (\ref{sf_RFfp_ta3}) provides more and more accurate approximations of the $\RFfp_{t}$ lines of the slip-forth mapping with respect to its distortion by footpoint advection.

Using equations (\ref{sbm_c}), (\ref{sbm_sh}), and (\ref{sbm}), a similar sequence of approximations can be found for the RFs of the slip-back mapping.
In the case of a negligible footpoint advection, one obtains that the $\RFfp$ lines of the instantaneous and past RFs, respectively, are given by
%
\begin{eqnarray}
  \RFfp^{\pm}_{t} & \approx & \QSSfp^{\pm}_{t} , 
	  \label{sb_RFfp_t}  \\
	\RFfp^{\pm}_{0} & \approx & \Pi_{t}(\QSSfp^{\mp}_{0}) .
	  \label{sb_RFfp_0a1}
\end{eqnarray}
If footpoint advection is not negligible but capable of producing only a moderate distortion of the slip-back mapping, the $\RFfp$ lines of the past RFs should be determined by
%
\begin{eqnarray}
	\RFfp^{\pm}_{0} & \approx & \Pi_{t} \circ F_{t} (\QSSfp^{\mp}_{0}) .
	  \label{sb_RFfp_0a2}
\end{eqnarray}
The presence of substantial interlocking or unlocking shears in the boundary motion requires, however, even a more accurate approximation, namely,
%
\begin{eqnarray}
	\RFfp^{\pm}_{0} & \approx & \Pi_{t}(\widetilde{\QSSfp}^{\mp}_{t}) ,
	  \label{sb_RFfp_0a3}
\end{eqnarray}
where $\widetilde{\QSSfp}^{\mp}_{t}$ denotes the $\QSSfp$ lines of the $\Pi^{\mathrm{ie}}_{t}$ mapping in the corresponding polarities.
With time running backward, the $\RFfp^{\pm}_{t}$ lines of the instantaneous RF will remain practically unchanged [see equation (\ref{sb_RFfp_t})], while the $\RFfp^{\pm}_{0}$ lines of the past RF will change and move on the boundary by sweeping the footprints of the flux tubes that have been reconnected within the time $t$ under study.

The evolving topology of these $\RFfp$ lines and footprints may be rather non-trivial, since the reconnection process may generally occur at different moments and places in the configuration by causing changes in its structure.
Such an intimate link of reconnection to the magnetic structure is expressed by the above approximate relationships between the $\RFfp$ and $\QSSfp$ lines.
The accuracy of these relationships can always be proved by using exact expressions (\ref{Qsf}) and (\ref{Qsb}) for the slip-squashing factors.
But even using these exact expressions, the reconnected or to-be-reconnected magnetic fluxes may be estimated only approximately.
The reason for this is that, in general, there is no strict difference between reconnection and magnetic diffusion, because the diffusion is always a part of the reconnection process.
In our formulation, this fact manifests itself in an inherent uncertainty of the end points of the unclosed $\RFfp$ lines.
Such end points can be identified by some threshold value $Q_{\mathrm{thresh}}$ of the slip-squashing factors.
Physically, this value is the degree of squashing of the elemental flux tubes that start at the end points of the $\RFfp$ lines and pass through the border of the reconnection region.
Since the location of this border is defined up to a fraction of the thickness of the reconnecting current layer, the threshold values $Q_{\mathrm{thresh}}$ and the end points of the unclosed $\RFfp$ lines also cannot be defined exactly.

From the physical point of view, the most accurate method for estimating them seems to rely on the voltage drop at the initial and final moments along the magnetic field lines that belong to the paths of the slip mappings starting at the respective $\RFfp$ lines.
Analyzing the distribution of this voltage drop along the $\RFfp$ lines would give us the most precise criterion for the end points of these lines---they would be simply the points where the voltage drop decreases to a fixed fraction of its nearest maximum at the $\RFfp$ line. 
However, such an approach requires additional computations involving additional data on electric fields in a given magnetic evolution, which seems to be hardly justifiable from the practical point of view.
The final result obtained by this electric field method will still not be unique but dependent on the threshold value of the voltage drop that is selected.
In this situation, it seems to be more practical to choose some large value of $Q_{\mathrm{thresh}}$ for identifying the end points of the $\RFfp$ lines by keeping in mind that the final result depends on the chosen $Q_{\mathrm{thresh}}$.
It should be emphasized that such an ambiguity of our method appears only in the case of the unclosed parts of the $\RFfp$ lines---their closed parts provide an unambiguous description of the reconnection process (see an example in section \ref{s:exmpl1}).
{{\hl
It is interesting that the closed $\RFfp$ lines seem to appear less often than the open ones.

There may also be the cases, where the transverse displacements of the $\RFfp$ lines are negligible, so that the $\RFfp$ lines evolve mainly by changing their length (see an example in section \ref{s:exmpl2}).
The reconnecting magnetic fluxes in such cases should be determined via the corresponding RF footprints defined at the boundary as areas with the slip-squashing factors larger than $Q_{\mathrm{thresh}}$.
Thus, this generalized definition of reconnecting flux tubes also does depend on $Q_{\mathrm{thresh}}$, which reflects an inherent property of the reconnection process in general 3D configurations.
}}
%

%
\section{EXAMPLE 1. FORMATION AND RISE OF A FLUX ROPE  \label{s:exmpl1}}

Let us illustrate now our theory of reconnection by considering a particular example of magnetic field evolution.
For this purpose, we choose the evolving configuration analyzed previously by \citet{Hesse2005} with the help of the GMR theory, whose major element is a nonvanishing voltage drop along the reconnecting magnetic field lines.
The consideration of this example will significantly facilitate a comparison between the two theories. 

In a Cartesian system of coordinates $(x,y,z)$, the corresponding components of the magnetic field are given by
%
\begin{eqnarray}
  B_{x} &=& -1 -  \frac{\varepsilon(t) (1-z^2/L_{z}^{2})}{(1+y^{2}/L_{y}^{2}) (1+z^2/L_{z}^{2})^{2}} , 
	\label{Bx} \\
  B_{y} &=& 0.2 , 
	\label{By}\\
  B_{z} &=& x , 
	\label{Bz}
\end{eqnarray}
where $L_{y}$ and $L_{z}$ are length-scale parameters, and $\varepsilon(t)$ is a monotonically growing function of time $t$. 
In contrast to \citet{Hesse2005}, the variables $y$ and $z$ are swapped, so that the plane $z=0$ corresponds to the photosphere.
Also the power degree 2 is recovered in the denominator of the second term in $B_{x}$ to correct a misprint in the original formula.
Only the $B_{x}$ component evolves with time via the function $\varepsilon(t)>0$, whose time derivative $\dot{\varepsilon}$ is assumed here to be strictly positive.
This implies the presence of an electric field with a nonvanishing component
%
\begin{eqnarray}
   E_{y} = - \frac{\dot{\varepsilon}(t) \, z}{(1+y^{2}/L_{y}^{2}) (1+z^{2}/L_{z}^{2})}  
	\label{Ey}
\end{eqnarray}
such that Faraday's law for the evolving $B_{x}$ is satisfied.

By construction, $B_{y}$ is nonvanishing and constant, so that the magnetic field (\ref{Bx})--(\ref{Bz}) has no null points. 
It is easy to prove also that, for the assumed $\varepsilon(t)>0$, $B_{x} < 0$ at the photospheric polarity inversion line defined by $x=z=0$.
Therefore, the transverse field at this line is always directed from the positive to negative polarity, which, in turn, means \citep{Titov1993} that this configuration has no coronal field lines touching the photosphere.
Thus, under these assumptions, the topology of magnetic field lines remains trivial during the evolution of the configuration.

\begin{figure}[htbp]
\epsscale{0.8}
\plotone{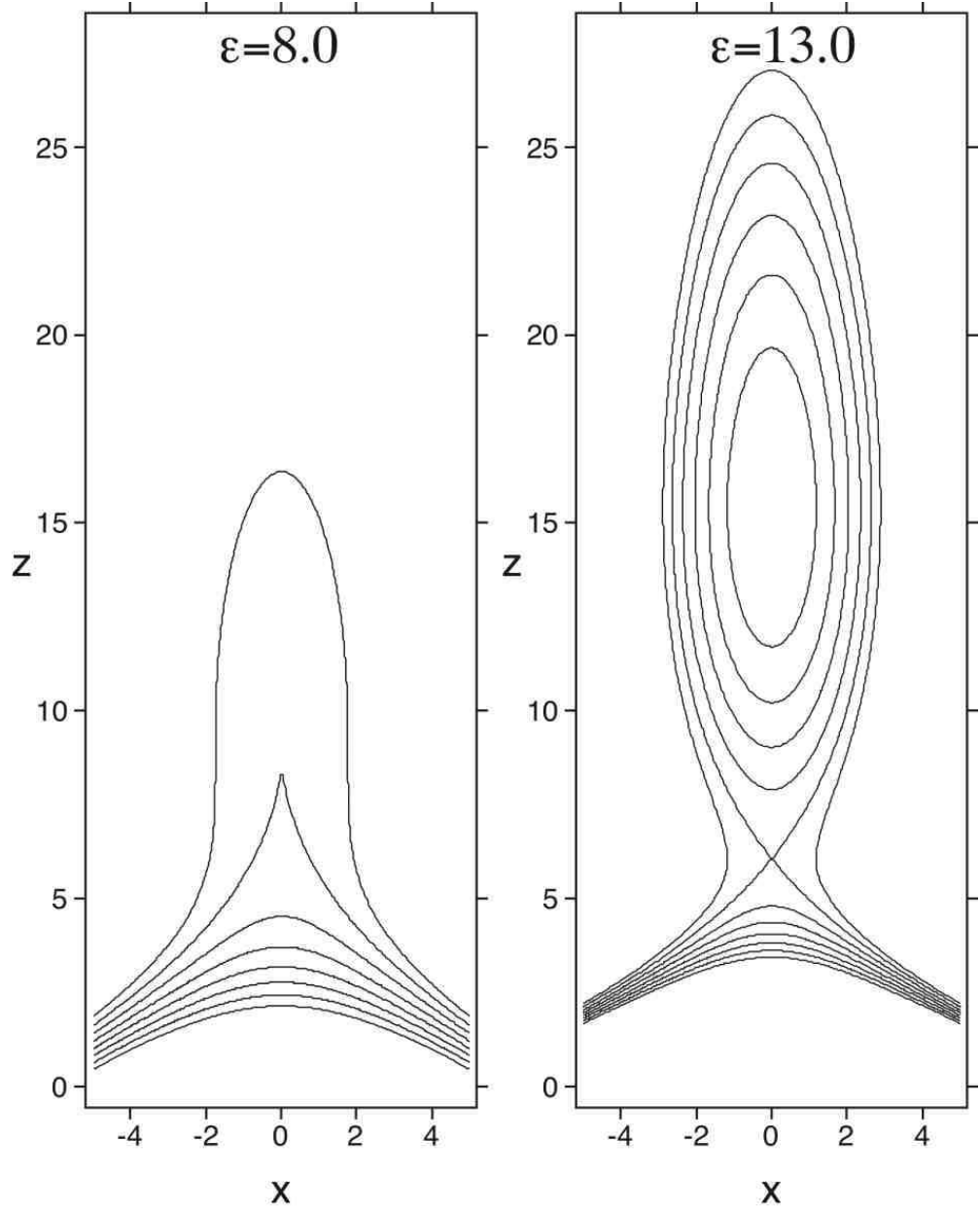}
\caption{Field lines of the transverse magnetic field $(B_{x},B_{z})$ in the middle plane $y=0$ at two different instances of the evolving magnetic configuration defined by equations (\ref{Bx})--(\ref{Bz}).
	\label{f:f4} }
\end{figure}

Nevertheless, its magnetic structure changes dramatically, which can be immediately seen from the evolution of the transverse magnetic field $\bm{B}_{\mathrm{trans}} \equiv (B_{x},B_{z})$ in the middle plane $y=0$.
Note that the field lines of $\bm{B}_{\mathrm{trans}}$ in this plane are contours of the function
\begin{eqnarray}
  f(x,z,t) = \frac{x^{2}}{2} + z \left( 1 + \frac{\varepsilon(t)}{1 + z^{2}/L^{2}_{z}} \right) .
	\label{f}
\end{eqnarray}
Figure \ref{f:f4} shows several such field lines at two different instances corresponding to $\varepsilon = 8 \equiv \varepsilon_{1}$ and $\varepsilon = 13 \equiv \varepsilon_{2}$.
They clearly demonstrate the formation of a magnetic island from the initial arcade-like configuration within the respective time interval.
This 2D plot provides, of course, only a first glimpse of the evolution of the magnetic structure, which is in fact fully three-dimensional, since the vector field $\bm{B}$ depends on all three coordinates.
This suggests, however, that the minimum points of $B^2$ might help to characterize the evolving magnetic structure.
In this approach, the formation of the magnetic island in the 2D plane can be viewed as a bifurcation of the cusp-type minimum into elliptic and hyperbolic minima.
This point of view is confirmed by a closer inspection of equations (\ref{Bx})--(\ref{By}).

\begin{figure}[htbp]
\epsscale{0.4}
\plotone{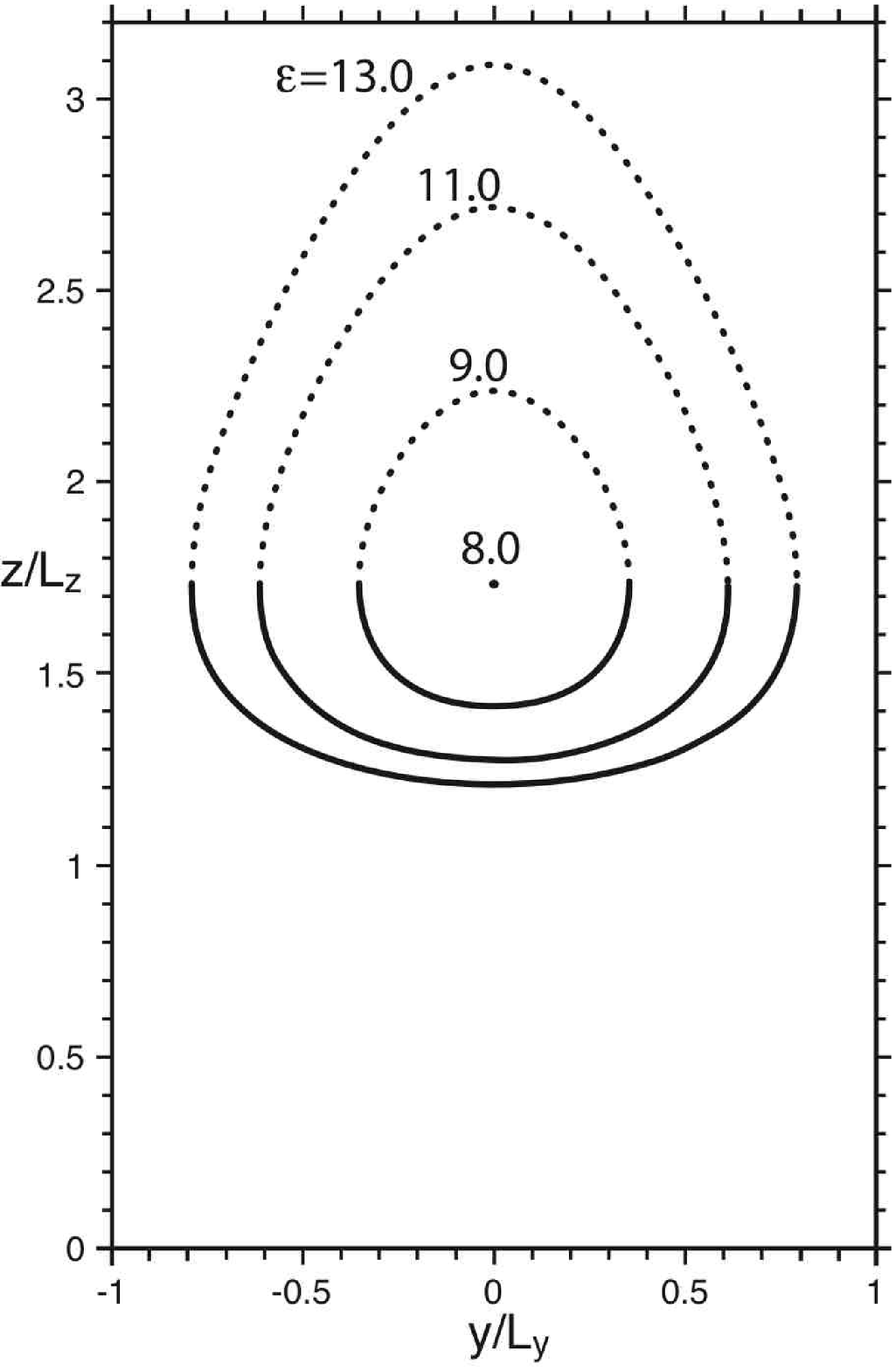}
\caption{Loci of minima of $B^{2}$ at four different times in the plane $x=0$. The solid and dotted lines, respectively, represent the minima of hyperbolic and elliptic types which emerge with growing $\varepsilon$ from the initial cusp point, present at $\varepsilon=8$.
	\label{f:f5} }
\end{figure}

Indeed, one can see from them that the minima of $B^2$ coincide with the nulls of $B_{x}$, whose locations are determined explicitly by 
\begin{eqnarray}
  \frac{z_{\mathrm{O},\mathrm{X}}}{L_{z}} =
     \left\{
       \pm \left[ \frac{2\varepsilon}{1+y^2/L_{y}^2}
             \left( \frac{\varepsilon}{8(1+y^2/L_{y}^2)} -1 \right) \right]^{1/2}
        + \frac{\varepsilon}{2(1+y^2/L_{y}^2)} - 1  
     \right\}^{1/2} .
	\label{zOX}
\end{eqnarray}
The upper and lower signs at the inner square root of this expression correspond to the loci of elliptic and hyperbolic minima, respectively, with heights $z_{\mathrm{O}}$ and $z_{\mathrm{X}}$.
The heights have real values only if $\varepsilon \ge 8$ and $|y| < y_{\mathrm{max}}$, where
\begin{eqnarray}
  \frac{y_{\mathrm{max}}}{L_{y}} = \left( \frac{\varepsilon}{8} - 1 \right)^{1/2}.
\end{eqnarray}
As $\varepsilon$ increases from $\varepsilon_{1} = 8$, a single minimum point of cusp-type $(y_{C}=0,\, z_{C}=\sqrt{3} L_{z})$ bifurcates into two curves of elliptic and hyperbolic minima (see Figure \ref{f:f5}).
It will be clear soon that some of these minima play a key role in the nonideal evolution of this configuration.

\begin{figure}[htbp]
\epsscale{0.9}
\plotone{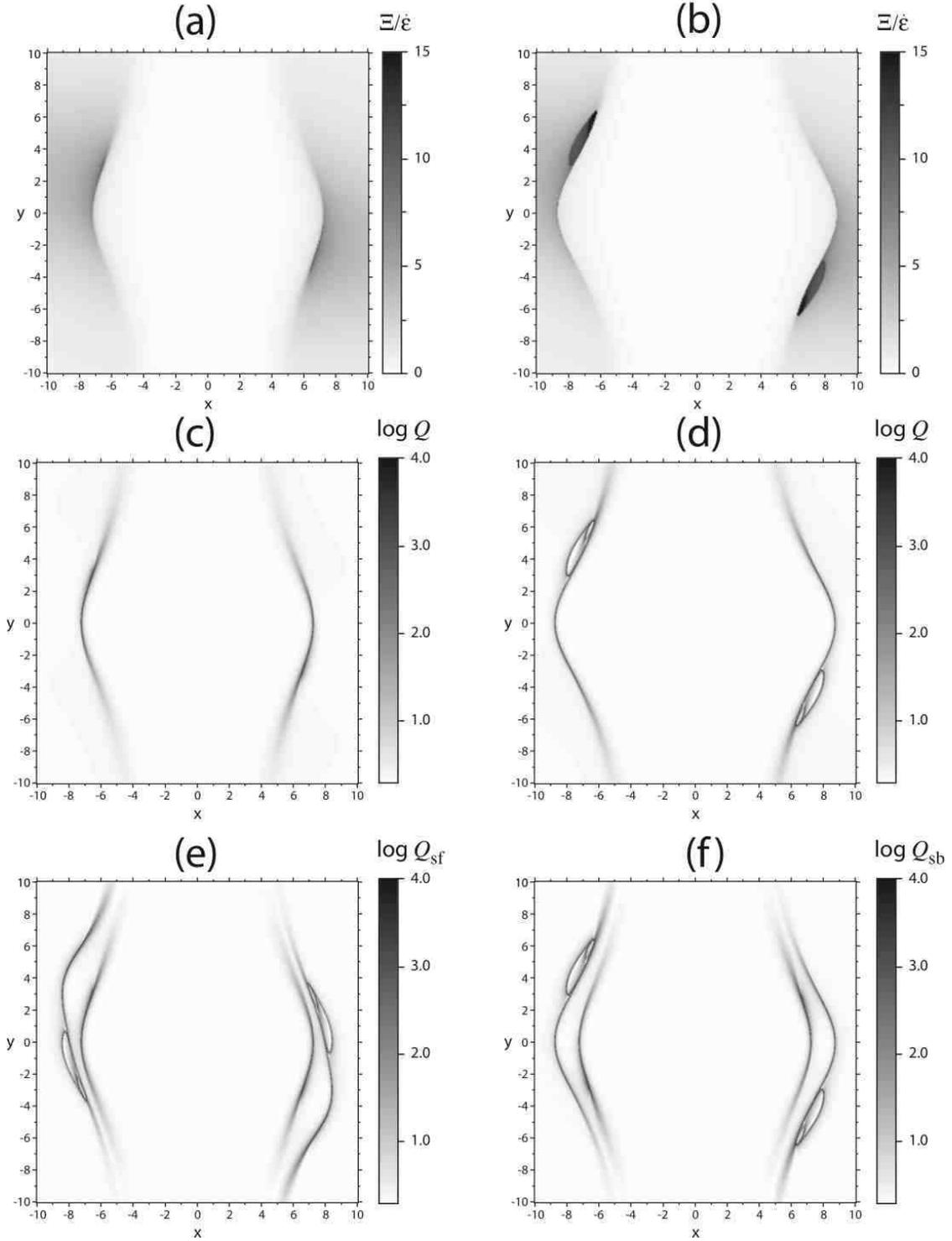}
\caption{Photospheric distributions of pseudopotential $\Xi$ (normalized to $\dot{\varepsilon}(t)$) [(a), (b)], squashing factor $Q$ [(c), (d)] at $\epsilon=8$ and $\epsilon=13$ (left and right panels, respectively) and slip-forth and slip-back squashing factors $Q_{\mathrm{sf}}$ and $Q_{\mathrm{sb}}$, respectively, between these two instants.
	\label{f:f6} }
\end{figure}

In a highly conducting plasma, the perpendicular component of an electric field ${\bm E}_{\perp} = E_{y} (\hat{\bm y} - B_{y} {\bm B}/B^{2})$ mainly sustains the advection of plasma elements with a drift velocity ${\bm v} = {\bm E}_{\perp} \times {\bm B}/B^2$.
In our example, this component vanishes at the photosphere ($z=0$) together with the total electric field $E_{y}$  [see equation (\ref{Ey})], which implies that the photospheric plasma elements remain at rest during the evolution of the configuration.
It is of particular importance that, in the volume, there is a nonvanishing parallel electric field ${\bm E}_{\|} =E_{y} B_{y} {\bm B}/B^{2}$, whose presence in the configuration provides a slippage of the footpoints with respect to the stationary photospheric plasma elements.
The latter was clearly demonstrated by \citet{Hesse2005}, who have also shown that such a slippage is most intensive at the magnetic field lines with the highest values of the voltage drop, called by them pseudopotential.
The latter is defined by
%
\begin{eqnarray}
  \Xi = - \int E_{\|} \, \mathrm{d}l ,
  \label{Xi}
\end{eqnarray}
where $l$ is an arc length of the magnetic field lines along which $E_{\|}$ is integrated between their end points, so that $\Xi$ is a function of these points.
For convenience, we have reproduced the photospheric distribution of $\Xi$, the same as in \citep{Hesse2005}, except that, in our case, $E_{\|}$ was integrated along the entire magnetic field lines without a cutoff by the lateral and top boundaries of numerical box.
This was done to eliminate the artificial distortion of the $\Xi$ distribution due to such a boundary effect.
The resulting $\Xi$ distributions at $\varepsilon=\varepsilon_{1}$ and $\varepsilon=\varepsilon_{2}$ are shown, respectively, in panels (a) and (b) of Figure \ref{f:f6}.
The second distribution differs only slightly from that calculated by \citet{Hesse2005}, which means that the mentioned boundary effect is not large for the chosen size of the numerical box.

For the same values of $\varepsilon$, panels (c) and (d) in Figure \ref{f:f6} present the photospheric distributions of the $Q$ factor in a logarithmic scale with the gray shading saturated at $\log Q = 4.0$.
Depending on the value of $\varepsilon$, the actual maxima of $\log Q$ in these distributions are two-three times larger than the saturation level. 
So the largest values of $Q$ are concentrated here in extremely narrow ridges tracing the footprints of the corresponding QSSs.
Comparison between the $Q$ and $\Xi$ distributions in Figure \ref{f:f6} [panels (c) and (d) vs. (a) and (b)] shows that the footprints of the QSSs pass exactly along the lines of the steepest gradient of $\Xi$, or, in other words, along the lines where the spatial variation of the footpoint slippage is the highest.
This indicates that the respective QSLs are indeed instantaneous RFs, as inferred in section \ref{s:rec_ft} from the general argument.
Further confirmation of this fact is obtained by considering the corresponding distributions of the slip-squashing factors.

\begin{figure}[htbp]
\epsscale{0.9}
\plotone{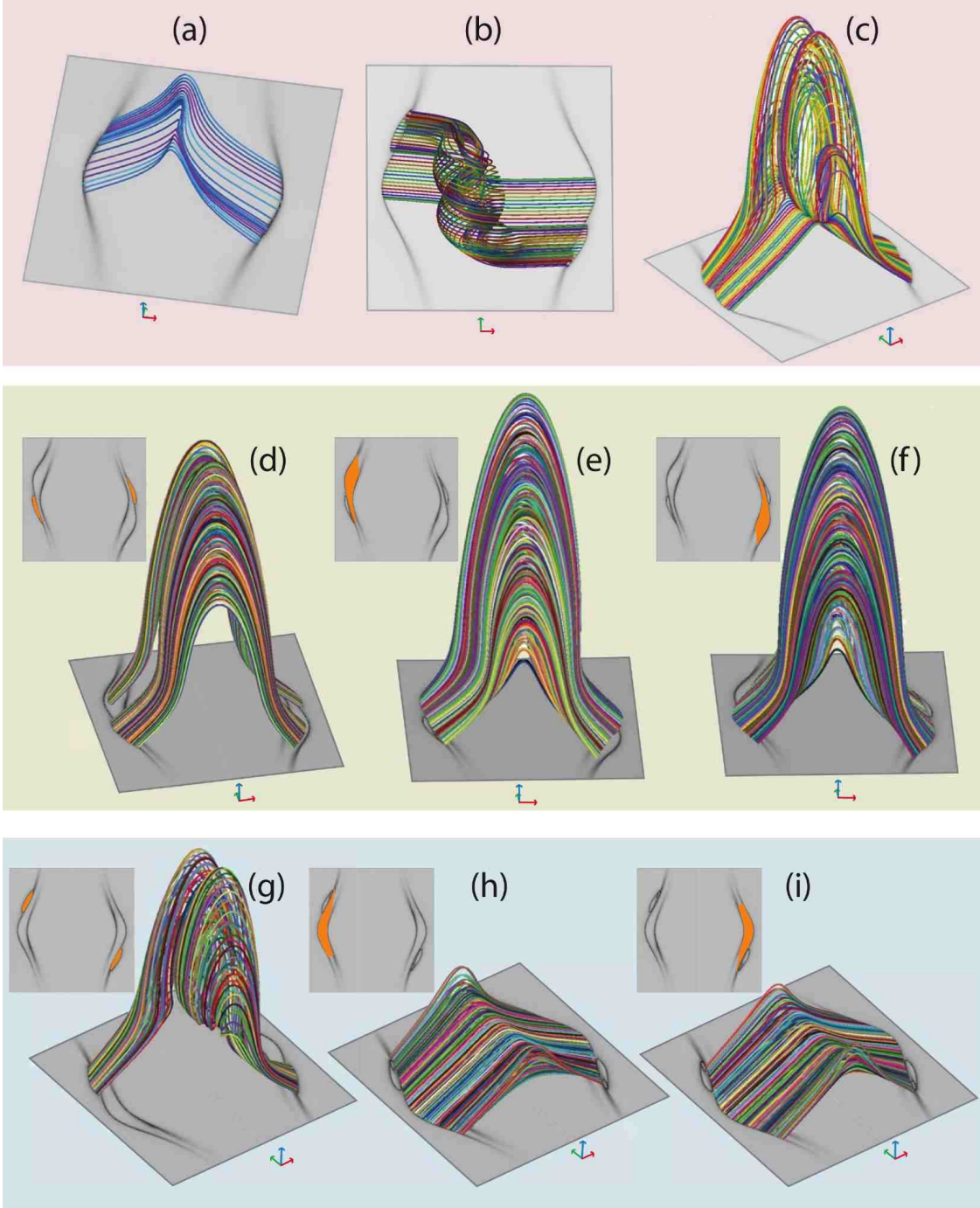}
\caption{Magnetic field lines passing in the vicinity of the minimum points of the cusp [(a), $\varepsilon=8$] and hyperbolic types [(b) and (c), $\varepsilon=13$] and the respective photospheric $Q$ distributions. To-be-reconnected [(d)--(f)] and reconnected [(g)--(i)] magnetic flux tubes at the initial ($\varepsilon=8$) and final ($\varepsilon=13$) moments, respectively; their launch footprints are shaded in orange and superimposed on the corresponding photosperic $Q_{\mathrm{sf}}$ and $Q_{\mathrm{sb}}$ distributions.  The respective orientation of the coordinate axes are shown by red ($x$), green ($y$) and blue ($z$) arrows.
	\label{f:f7} }
\end{figure}

Before starting such a consideration, let us look first at the structure of the magnetic field lines in the vicinity of the QSSs.
Figure \ref{f:f7}a shows a set of the field lines at $\varepsilon=\varepsilon_{1}$ with the launch points chosen to be very close to the cusp minimum point of $B^{2}$.
It is clearly seen there that the footpoints of these field lines are distributed exactly along the ridges of the $Q$ distribution.
So such field lines approximate the QSS by showing that it has a wedge-like shape with the sharpest corner at the cusp minimum point.
Panels (b) and (c) in Figure \ref{f:f7} depict a set of field lines at $\varepsilon=\varepsilon_{2}$ with the launch points near the loci of the hyperbolic minima of $B^{2}$ defined by equation (\ref{zOX}).
The footpoints of these field lines are also distributed exactly along the ridges of the corresponding $Q$ distribution.
So such field lines approximate the QSS by demonstrating that it consists of two parts attached to each other.
The lower part has a wedge-like shape with a sharp corner along the loci of the hyperbolic minima.
The upper part is a tube-like surface enclosing a twisted flux rope that lies on the top of this wedge.
It starts at the photosphere as a narrow, very flat, and almost untwisted flux tube spreading along the slopes of the wedge and blows up near the wedge corner into a very large, elongated, and highly twisted plasmoid.
The cross section of the resulting QSS by the middle plane $y=0$ coincides with the separatrix line of the transverse magnetic field shown in the right panel of Figure \ref{f:f4}.
Thus, the QSLs in our example are due to the presence of the hyperbolic or cusp minima of $B^{2}$---this fact will be explained in section \ref{s:min} from the general point of view. 

{{\hl
In the strict sense, the oval-like parts of the $\QSSfp$ lines (Fig. \ref{f:f6}d) are not closed but open lines---the ridges of the $Q$ distribution here approach very close to one another when making their turns but they never completely merge.
We believe that this was also the case in the twisted configuration analyzed previously by \citet{Demoulin1996a} with the help of the norm $N$ rather than $Q$.
Such a behavior of QSLs in both cases has a simple explanation: due to a relatively large twist of the rope, there are field lines with a different number of windings (larger than one) in the neighborhood of the hyperbolic magnetic minima in the corona.
A relative proximity of such field lines to these minima is translated into the apparent closeness of the respective $\QSSfp$ lines at the photosphere due to an enormous squashing that the elemental flux tubes experience on their way from the minima to the photosphere.
From the observational point of view, however, this seems to be a atypical situation, since the observed flaring bright loops rarely show multiple windings.
Therefore, the QSL footprints in realistic magnetic configurations should look like open rather than closed curves, as actually demonstrated by \citet{Titov2007a} for a flux rope configuration with a realistic amount of twist.
}}
%

Consider now the distributions of the slip-forth and slip-back squashing factors calculated numerically for the variation of $\varepsilon$ in the interval $[\varepsilon_{1}, \varepsilon_{2}]$.
These calculations are significantly simplified due to the above-mentioned line-tying condition $\left. {\bm E}\right|_{z=0}=0$ that implies the absence of plasma flows at the photosphere.
As discussed in the previous section, the calculation of $Q_{\mathrm{sf}}$ or $Q_{\mathrm{sb}}$ in such a case requires knowledge only of ${\bm B}$ at the initial and final moments, which can be obtained here from equations (\ref{Bx})--(\ref{Bz}).
The resulting distributions of $Q_{\mathrm{sf}}$ and $Q_{\mathrm{sb}}$ are shown in panels (e) and (f), respectively, in Figure \ref{f:f6}.
The inner or outer pairs of the ridges of the $Q_{\mathrm{sf}}$ distribution trace the instantaneous or future $\RFfp$ lines, respectively;
and vice versa: the outer or inner pairs of the ridges of the $Q_{\mathrm{sb}}$ distribution trace the instantaneous or past $\RFfp$ lines, respectively.
Comparing panels (e) and (f) vs. (c) and (d) in Figure \ref{f:f6} shows that the instantaneous $\RFfp$ lines practically coincide with the instantaneous $\QSSfp$ lines---exactly as previously inferred from the general consideration [see equations (\ref{sf_RFfp_0}) and (\ref{sb_RFfp_t})].
In contrast, the future and past $\RFfp$ lines  are noticeably distorted in comparison with the corresponding $\QSSfp$ lines, which is also in accord with the inferred general equations (\ref{sf_RFfp_ta1}) and (\ref{sb_RFfp_0a1}).
{{\hl
Such a distortion is sufficiently smooth, that it causes the oval-like parts of the corresponding RF footprints to appear nearly as closed as their corresponding QSL footprints.
}}
%

%
It is useful also to consider the behavior of the $\RFfp$ lines in response to the variation of $\varepsilon_{2}$ or $\varepsilon_{1}$ for the slip-forth or slip-back mappings, respectively.
In both cases, the footprints of the instantaneous $\RFfp$ lines remain practically unchanged under such a variation by recurring the shape of the corresponding $\QSSfp$ lines shown in panels (c) and (d) in Figure \ref{f:f6}.
On the contrary, the future and past $\RFfp$ lines change their locations, shapes, and even topology.

Let $\varepsilon_{1}=8$ be kept constant and $\varepsilon_{2}$ be increasing in value from 8 to 13.
Then the future $\RFfp$ line in each of the magnetic polarity will split and move outward from the respective instantaneous $\RFfp$ line by gradually increasing its length and bending itself in the middle part.
In addition to these transformations, two small and very flat ovals emerge in the future $\RFfp$ lines by stretching themselves out from the conjugate footpoints of the field line that passes in the volume through the cusp minimum point.
Such ovals enclose the footprints of the initial flux tubes that will reconnect at the moment $\varepsilon_{2}$ and form the above-mentioned flux rope---these flux tubes are shown in Figure \ref{f:f7}d.
The areas swept out by the major parts of the future $\RFfp$ lines determine the footprints of the flux tubes that will reconnect and form the ``flare loops" below the flux rope.
One can see from Figures \ref{f:f7}e and \ref{f:f7}f that, as expected, their conjugate footprints also include the magnetic flux that will belong to the flux rope.
Finally, the areas that are swept by the ovals in the exterior determine the footprints of the flux tubes that will pass through the flux rope formed during this evolution.

Let $\varepsilon_{2}=13$ be kept constant and $\varepsilon_{1}$ be decreasing in value from 13 to 8.
Then the past $\RFfp$ line in each of the magnetic polarity will split and move inward from the respective instantaneous $\RFfp$ line by gradually decreasing its length and contracting the ovals that enclose the footprints of the flux rope (see Figure \ref{f:f7}g).
At the moment $\varepsilon_{1}=8$, the ovals will completely shrink to the conjugate footpoints of the field line that passes in the volume through the cusp minimum point.
The areas swept out by the major parts of the past $\RFfp$ lines determine the footprints of the reconnected flux tubes manifesting themselves as ``flare loops" formed below the flux rope [panels (h) and (i) in Figure \ref{f:f7}].
{{\hl
This kind of flux tube may have been directly observed in X-rays as coronal loops slipping along their respective QSL footprints \citep{Aulanier2007}.
}}
Finally, the areas that are swept by the ovals in the exterior determine the footprints of the flux tubes that have passed through the flux rope formed during this evolution.

Thus, this example clearly demonstrates that the slip-squashing factors allow one to identify reconnecting magnetic fluxes in a given magnetic evolution with an unprecedented level of detail.

%
{{\hl
\section{EXAMPLE 2. RECONNECTION IN A CURRENT-LAYER PATCH
	\label{s:exmpl2}}
Let us consider now an alternate magnetic evolution, where, in contrast to Example 1, the footprints of the initial and final QSLs overlap each other.
This causes the resulting RFs to behave in a less predictable way than before, and it also provides some new insights into the significance of the slip-squashing factors.

\begin{figure}[htbp]
\epsscale{1.0}
\plotone{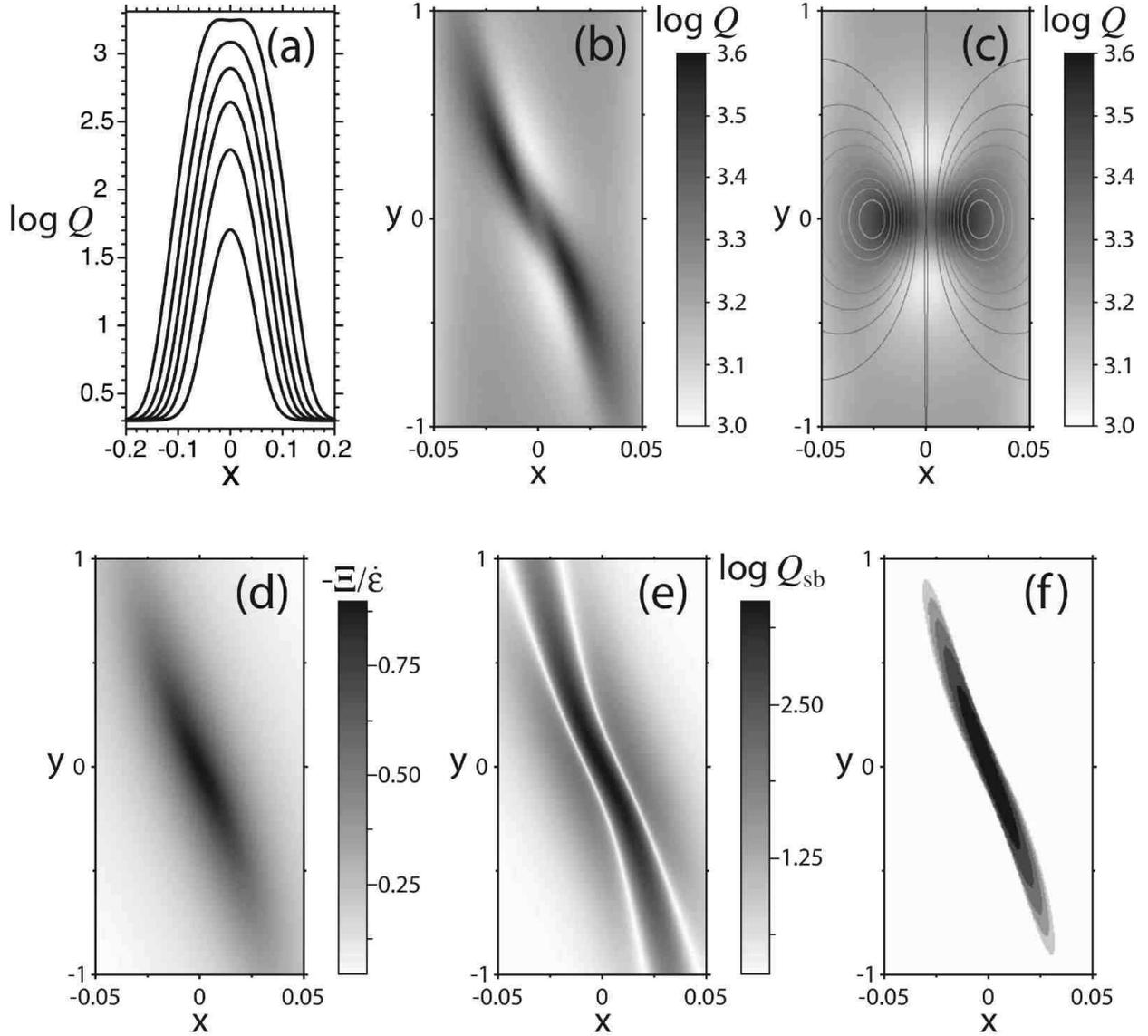}
\caption{
Profiles of $Q$ across the initial slab configuration ($\varepsilon=0$, $l_{\rm sh}=0.05$, $L_{x}=0.025$ and $L_{y}=L_{z}=0.3$) at  $\theta_{0}=k\,\pi/18$, with integers $k=1,\ldots, 6$ (a).  The distributions of $Q$ at $z=-1$ (b) and $z=0$ (c), the distributions of $-\Xi/\dot{\varepsilon}$ (d) and $Q_{\rm sb}$ (e) at $z=-1$ in the perturbed slab configuration with $\varepsilon=0.006$ and $\theta_{0}=\pi/3$ (the remaining parameters are the same); the contours of $B^{2} = 0.91+0.01\,k$,  $k=0,\ldots,9$, are superimposed on the respective $Q$ distribution in panel (c).  Panel (f) shows for the same parameters the gray-shaded areas in the plane $z=-1$ where $\log Q_{\rm sb} \ge 2$ for four different times corresponding to $\varepsilon=0.001\,k$, $k=2, \ldots, 6$, with a stepwise decrease of shading from the initial to the final times.
	\label{f:f8} }
\end{figure}

As the initial configuration, let us take the force-free magnetic field depending only on the $x$ coordinate as
\begin{eqnarray}
  {\bm B}_{0} &=& \sin\theta(x) \hat{\bm y} + \cos\theta(x) \hat{\bm z} ,
	\label{B0_sl} \\
	\theta(x) &=& \theta_{0} \tanh(x/l_{\rm sh}) .
	\label{th(x)}
\end{eqnarray} 
The corresponding magnetic field lines are straight and tilted toward the $z$-axis at the angle $\theta(x)$, which changes smoothly with $x$ from $-\theta_{0}$ to $\theta_{0}$ to form a sheared structure. 
Most of the magnetic shear, and its associated electric current density,
\begin{eqnarray}
  {\bm j} = \theta^{\prime}(x) {\bm B}_{0} =\frac{\theta_{0}/l_{\rm sh}}{\cosh^{2}(x/l_{\rm sh})} {\bm B}_{0} 
	\label{j_sl}
\end{eqnarray}
are concentrated in a slab-like layer of half thickness $l_{\rm sh}$.
Let this configuration be bounded from the top and bottom by ideally conducting plates located at $z=\pm L \equiv \pm 1$.
Then using the straightness of the magnetic field lines and equations (\ref{D})-(\ref{G}), it is not difficult to calculate the respective squashing factor
\begin{eqnarray}
  Q = 2 + \frac{4\theta_{0}^{2} L^{2}/l_{\rm sh}^{2}}{\cosh^{4}(x/l_{\rm sh}) \, 
       \cos^{4}\theta(x)} .
	\label{Q_sl}
\end{eqnarray}
Figure \ref{f:f8}(a) shows the profiles of $Q$ at different $\theta_{0}$ and other fixed parameters by demonstrating that at sufficiently small $l_{\rm sh}$ and large $\theta_{0}$ our current layer becomes a QSL, in which both $Q$ and current density increase together with the total shear angle $2 \theta_{0}$.
Since $\left|{\bm B}_{0}\right|=1$ everywhere in the configuration, the appearance of such a QSL is not related to the presence of any magnetic minima in contrast to Example 1.
Instead, this particular QSL is the result of a large magnetic shear present inside a thin layer with a strong current density.

If the latter reaches a certain threshold value defined through other plasma parameters, a nonideal plasma effect may come into play.
For example, the onset of some plasma instability might be triggered.
This would, in turn, cause the appearance of an electric field with a nonvanishing component along ${\bm B}_{0}$ inside the layer.
One can assume also that in reality some of the plasma parameters may have nonuniform distributions and so the threshold conditions would be reached first at some spot in the layer.
This would, in turn, lead to a localization of the parallel electric field in the spot.
We can model such a process in our case by introducing into the configuration the following electric field
\begin{eqnarray}
  {\bm E} = \dot{\varepsilon}(t) f(x,y,z) \hat{\bm z} .
	\label{E_sl} 
\end{eqnarray}
The time and space variation of ${\bm E}$ is described here by the time derivative of some function $\varepsilon(t)$ and the function
\begin{eqnarray}
	  f(x,y,z) = \left[x^{2}/L_{x}^2+y^{2}/L_{y}^2+\cosh(z/L_{z})\right]^{-1} ,
	\label{f2}
\end{eqnarray}
respectively.
Taking the curl of ${\bm E}$ and integrating it over time, we obtain in accord with Faraday's law the resulting magnetic field 
\begin{eqnarray}
  {\bm B} = {\bm B}_{0} + \varepsilon \hat{\bm z} \times \bm{\nabla} f .
	\label{B_sl} 
\end{eqnarray}
This is, of course, a very approximate model of the respective real process, since ${\bm E}$ is prescribed here in an ad hoc way rather than derived from the appropriate plasma dynamics.
However, the model is accurate enough to qualitatively demonstrate the difference between magnetic reconnection and diffusion in a simple, but generic, case.

The field-line equation does not admit an analytical solution for the derived magnetic field ${\bm B}$, so the corresponding squashing factors have been calculated numerically for the following set of parameters: $\theta_{0}=\pi/3$, $l_{\rm sh}=0.05$, $L_{x}=0.025$, $L_{y}=L_{z}=0.3$, and $\varepsilon = 0.001\, k$ with integers $k=1 \ldots 6$.
In all these cases, the assumed electric field is localized in a patch-like region that has the shape of a very flat ellipsoid compressed in the $x$ direction.
Figure \ref{f:f8}(b) shows an example of the $Q$ distribution in the plane $z=-1$ at $\varepsilon=0.006$, when the transverse gradients of the perturbed and unperturbed $y$ components of the magnetic field become comparable in value at the center of the patch.
The respective $Q$ distribution at the top plate $z=+1$ is obtained from the one below by its mirror reflection about the $y$-axis.
These distributions show that two well-pronounced QSLs adjacent to one another eventually emerge inside the initial QSL.
Such an evolution of the QSLs is very different from the one studied in Example 1.

By definition, $Q$ is invariant to the direction of mapping along the field lines, so its footpoint value can be prescribed by any other point of the respective field line, thereby extending the boundary $Q$ distribution into the volume.
This allows one to study the 3D structure of the QSLs via the distribution of $Q$ at different cross sections.
Figure \ref{f:f8}(c) displays one such distribution in the plane $z=0$ at  $\varepsilon=0.006$ together with the respective set of contours of $B^2={\rm const}$ in this plane.
One can see from this figure that the positions of local maxima of $Q$ and minima of $B^2$ in the plane $z=0$ are relatively close to one another, although not identical.
A noticeable shift between them is probably the result of the relative shallowness of the minima of $B^2$, whose values are just 10\% lower in comparison with the background, while the characteristic sizes of the minima are of the order of $l_{\rm sh}$.
In other words, the appearance of the new QSLs seems to be related again to the formation of the minima of $B^2$ in the configuration, as was the case in Example 1.
The only difference is that such a relation in Example 1 is stronger because of a much larger depth of the corresponding minima.


For the same set of parameters, Figure \ref{f:f8}(d) shows the distribution of the pseudo-potential  $\Xi$ [defined by equation (\ref{Xi})] in the plane $z=-1$, which suggests that a substantial change of magnetic connectivity must occur in our configuration along the diagonal of the computational rectangular area.
This is confirmed by Figure \ref{f:f8}(e) that presents the respective distribution of the slip-back squashing factor $Q_{\rm sb}$ for the time period, where $\varepsilon$ increases from 0 to 0.006.
Since the assumed electric field drops exponentially with growing $|z|$ toward the boundary plates [see equations (\ref{E_sl})--(\ref{f2})], the $Q_{\rm sb}$ distribution is calculated here under the assumption that the line-tying condition is strictly fulfilled at the plates.
In other words, a possible small footpoint advection is neglected in calculating the respective slip-back mapping, which is composed of the final and initial field-line mappings only, as described in section \ref{s:nonideal}.
The $Q_{\rm sb}$ so obtained demonstrates that a major change of connectivity occurs along a narrow strip that is aligned with both the respective new QSL footprints and the area with large values of $|\Xi|$.
The latter is in good agreement with the analysis of a similar nonideal evolution in the so-called X-line configuration \citep{Hesse2005}.
The indicated narrow strip can be identified here only with the footprint of the instantaneous RF, since the past RF is absent at the initial moment (simply because the new QSLs just start to emerge at that moment).

To see how this RF footprint evolves, we determine it as the area at the boundary plate, where $Q_{\rm sb} \ge Q_{\rm thresh} = 10^2$, for several pairs of $\varepsilon$, whose initial values equal 0.
This value is used same for all the pairs, while the final values vary within the interval $[0,\: 0.006]$.
These RF footprints, represented in Figure \ref{f:f8}(f) by gray shadings of different intensity, form a clearly nested structure such that the area that is swept out by the final time approaches asymptotically to the area of the instantaneous RF footprint at this time. 
In terms of the corresponding $\RFfp$ lines, this means that they grow in length by remaining nearly self-aligned and not by moving in the transverse direction (in contrast to Example 1).
Therefore, in order to determine the reconnected flux tubes in situations like this, one should extend our previous definition of the reconnected boundary area by replacing the sweeping $\RFfp$ lines in this definition with the sweeping RF footprints.
Since both of them depend on the chosen value $Q_{\rm thresh}$, such an extension does not increase the ambiguity in determining the reconnected magnetic fluxes.
It does, however, make this definition more general.

\begin{figure}[htbp]
\epsscale{0.9}
\plotone{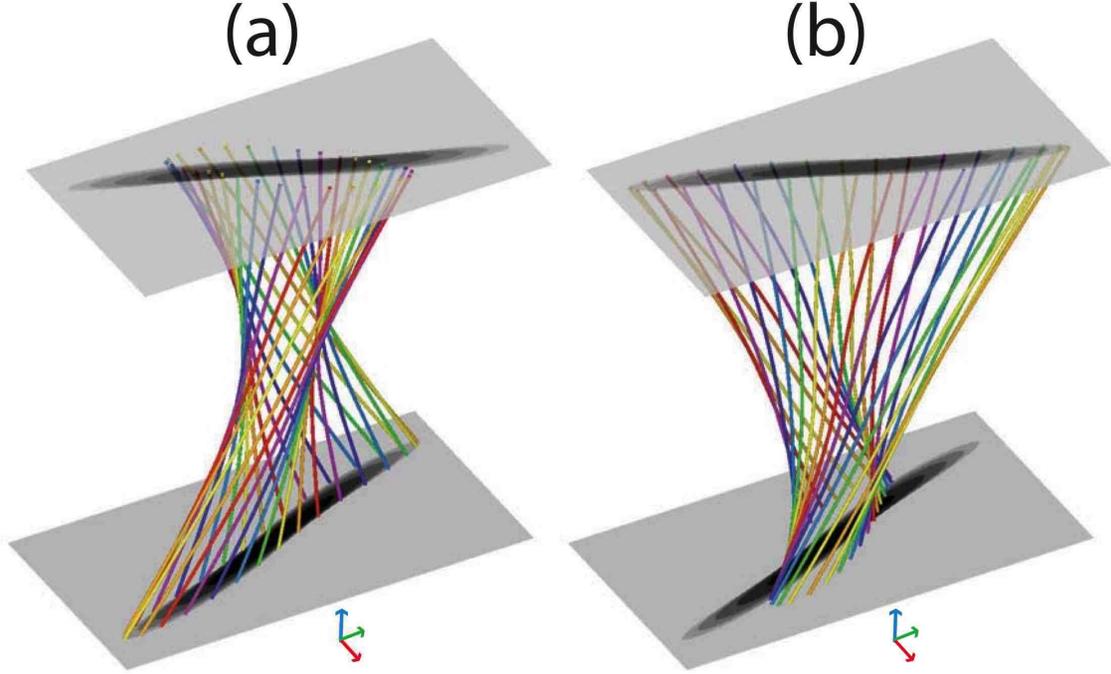}
\caption{
{\hl
Magnetic field lines in the slab configuration at $\varepsilon=0.006$ with other parameters, the same as in Figure \ref{f:f8}. The launch points are located in the planes $z=-1$ (a) and $z=+1$ (b) along the contours $\log Q_{\rm sb} = 2.0$. The respective orientation of the coordinate axes are shown by red ($x$), green ($y$), and blue ($z$) arrows.  The length scale in the $x$-direction is stretched ten times.
} 
	\label{f:f9} }
\end{figure}

The reconnected flux tubes as defined above are shown in Figure \ref{f:f9}  with the help of the magnetic field lines that are launched from the contours $\log Q_{\rm sb} = 2.0$ at the boundary plates $z=\pm1$.
These flux tubes intersect but do not coincide with one another, which is evident from the locations of the target footpoints being scattered partly outside of the RF footprint at the other boundary plate.
Nevertheless, such footpoints also outline a very elongated oval that is slightly turned with respect to the corresponding RF footprint around their common center $(x=0,\: y=0)$.
This is a manifestation of the characteristic plasma rotation in magnetic flux tubes passing through a nonideal region---an effect that was previously described in the framework of the GMR theory \citep{Schindler2006}.

Our second example suggests that the characteristic feature that distinguishes reconnection from diffusion is RFs with narrow ribbon-like footprints.
Such footprints are characterized by two length scales, one of which is much smaller than the other.
Displacement of plasma elements in the nonideal region across the corresponding RFs causes a large change in magnetic connectivity along the RF footprints.
In other words, the high values of slip-squashing factors at the flux tubes that pass through and in the vicinity of a nonideal region is the major signature of reconnection compared to diffusion.
The transition from one process to the other is not abrupt but continuous both in time and in parameter space, and only in this sense it is just a matter of semantics of how one refers to this process---reconnection or diffusion.
Generally, the appearance of such nonideal regions in configurations is closely related to locally enhanced current density, which corresponds to large magnetic shear, large values of the squashing factor, and hence large values of the slip-squashing factors.
Because of this causal chain, it seems preferable to refer to the process in the nonideal regions as reconnection rather than magnetic diffusion.
}}

Note also that strong outflow jets of plasma from the nonideal regions have been a feature of traditional 2D reconnection models.
We may conjecture that perhaps this is also a property of many 3D
reconnection regimes, although perhaps some types of 3D reconnection
do not possess rapid outflow jets [e.g. \citet{Schindler1988}].
The accurate proof of such a conjecture must rely on a self-consistent analysis of the full system of MHD equations, which is beyond the scope of the present work.
However, some indications of its validity can be seen from our present consideration as well.
In particular, equations (\ref{f2}) and (\ref{B_sl}) show that the considered nonideal process causes an X-type perturbation of the transverse magnetic field.
Therefore, the Lorentz force due to interaction of this perturbation with the initial strong current layer will tend to accelerate plasma out of the nonideal region and to produce outflow jets.
Thus, the thinner the current layer, the stronger the acceleration, which corresponds, in turn, to the narrower RFs developing in the configuration, as demonstrated in this section.


\section{MAGNETIC MINIMUM POINTS
   \label{s:min}}

Determining QSLs and RFs via computation of the corresponding squashing factors is conceptually simple but technically difficult.
{{\hl
However, the particular examples of magnetic evolution considered in sections \ref{s:exmpl1} and \ref{s:exmpl2} strongly suggest that there is a close relationship between QSLs, RFs, and magnetic minimum points.
If the presence of such a minimum point is a sufficient condition for the existence of the respective QSL, this might significantly simplify determining QSLs in a given configuration.
}}
Finding magnetic minima is technically as simple as finding magnetic null points, so locating such minima would significantly facilitate the localization of QSLs and RFs.
This motivates us to study the local properties of magnetic minima in the most general terms.

Note first that, for any point with $B^2\equiv B_{0}^{2} \ne 0$, the system of coordinates can always be chosen in such a way that $B_{x}=B_{y}=0$, while  $B_{z} \equiv B_{0}$.
If the point is a magnetic minimum, then $\nabla B^2 = \nabla B_{z}^{2} = 0$ or simply $\nabla B_{z}=0$ in such a system of coordinates.
Taking also into account that $\nabla\cdot {\bm B} = 0$, we obtain the following general expression for the matrix of gradients of magnetic field at its minimum point:
%
\begin{eqnarray}
  {\cal B} \equiv 
   \left[ \nabla\bm{B} \right] =
\left(
  \begin{array}{crc}
        \partial_{x}B_{x} & 
        \partial_{y}B_{x} &
        \partial_{z}B_{x}   \\ [3pt]
        \partial_{x}B_{y} & 
       -\partial_{x}B_{x} &
        \partial_{z}B_{y}   \\ [3pt]
        0 &  0 &  0      
  \end{array}
\right)
\equiv 
\left(
  \begin{array}{crc}
        {a} &    \tilde{b}    &  {c}   \\ [3pt]
        {b} &         -{a}    &  {d}   \\ [3pt]
         0  &           0     &   0      
  \end{array}
\right) .
	\label{BmM}
\end{eqnarray}
Rotating the system of coordinates about the $z$-axis does not change the components of ${\bm B}$ at the minimum point, but it reduces ${\cal B}$ at a certain rotation angle to the form
%
\begin{eqnarray}
  {\cal B}  = 
\left(
  \begin{array}{crc}
        a &  -b  &  c   \\ [3pt]
        b &  -a  &  d   \\ [3pt]
        0 &   0  &  0      
  \end{array}
\right) ,
	\label{BmMcan}
\end{eqnarray}
where all matrix elements are generally different from those in equation (\ref{BmM}).
Thus, in this rotated system of coordinates, we obtain the canonical linearized form for a magnetic-minimum field 
%
\begin{eqnarray}
  B_{x} & = & a x - b y + c z ,
	\label{Bxm}   \\
  B_{y} & = & b x - a y + d z ,
	\label{Bym}   \\
  B_{z} & = & B_{0}.
	\label{Bzm}
\end{eqnarray}
Using these expressions, one can easily find from equation (\ref{FLE}) the corresponding magnetic field lines.
Depending on the values of parameters $a$ and $b$, there are only three different types of minimum: hyperbolic, cusp, and elliptic minima, exactly as in the example considered in section \ref{s:exmpl1}.
Assuming that the field lines start in the plane $z=0$, so that $\left.(x,y,z)\right|_{\tau=0}=(x_{0}, y_{0}, 0)$, we have for all these types of minimum
\begin{eqnarray}
   z &=& B_{0}\, \tau ,
	  \label{zHCE}
\end{eqnarray}
which simply represents linear dependence of $z$ on the differential flux tube volume $\tau$.
For hyperbolic minima, which exist when $|a|>|b|$, the field lines are given by
%
\begin{eqnarray}
  x &=& x_{0} + (a x_{0}- b y_{0}) \frac{\sinh\mu\tau}{\mu} + c B_{0}\frac{\cosh \mu \tau - 1}{\mu^{2}}  + (ac-bd) B_{0} \frac{\sinh\mu\tau - \mu \tau}{\mu^{3}} ,
	  \label{xH}  \\
	y &=& y_{0} + (b x_{0}- a y_{0}) \frac{\sinh\mu\tau}{\mu} + d B_{0}\frac{\cosh \mu \tau - 1}{\mu^{2}}  + (bc-ad) B_{0} \frac{\sinh\mu\tau - \mu \tau}{\mu^{3}} ,
	  \label{yH} 
\end{eqnarray}
where
$
  \mu = \left( a^{2} - b^{2} \right)^{1/2} .
$
In the limit $\mu \rightarrow 0$ yielding $|a|=|b|$, one can retrieve from here the field-line structure at cusp minima described by
%
\begin{eqnarray}
  x &=& x_{0} + a (x_{0}-y_{0}) \tau + c B_{0} \frac{\tau^{2}}{2}
           + a(c-d) B_{0} \frac{\tau^{3}}{6} ,
	  \label{xC}  \\
	y &=& y_{0} + a (x_{0}-y_{0}) \tau+ d B_{0} \frac{\tau^{2}}{2}
           + a(c-d) B_{0} \frac{\tau^{3}}{6} .
	  \label{yC}
\end{eqnarray}
The same expressions can also be retrieved in the limit $\omega \rightarrow 0$ from 
the elliptic minima, which exist when $|a| < |b|$ with neighboring field lines determined by
%
\begin{eqnarray}
  x &=& x_{0} + (a x_{0}- b y_{0}) \frac{\sin \omega\tau}{\omega} + c B_{0}\frac{\cos \omega \tau - 1}{\omega^{2}}  + (ac-bd) B_{0} \frac{\sin \omega\tau - \omega \tau}{\omega^{3}} ,
	  \label{xE}  \\
	y &=& y_{0} + (b x_{0}- a y_{0}) \frac{\sin \omega\tau}{\omega} + d B_{0}\frac{\cos \omega \tau - 1}{\omega^{2}}  + (ad-bc) B_{0} \frac{\sin \omega\tau - \omega \tau}{\omega^{3}} ,
	  \label{yE}  
\end{eqnarray}
where
$
  \omega = \left( b^{2} - a^{2} \right)^{1/2} .
$

The above analytical expressions show that the highest divergence of field lines in response to the variation of $x_{0}$ or $y_{0}$ is achieved in the vicinity of hyperbolic magnetic minima [see equations (\ref{xH}) and (\ref{yH})].
The corresponding gradients of $x$ and $y$ with respect to $x_{0}$ or $y_{0}$ grow exponentially with $z$, and the rate of this growth increases with decreasing $B_{0}$.
In the case of cusp minima, however, such a growth is only linear in $z$, while for elliptic minima it is even saturated via the sine-function.
This provides a strong general argument in favor of the idea that the hyperbolic and cusp minima of a magnetic field are responsible for the appearance of QSLs and corresponding RFs in evolving magnetic configurations. 

{{\hl
Although both of the examples considered in sections \ref{s:exmpl1} and \ref{s:exmpl2} possess some symmetry, we do not think that this is a crucial factor for the emergence of QSLs and RFs.
The really important parameters that control the thickness of the emerging QSLs should be the depth of the respective hyperbolic magnetic minima and their characteristic sizes compared to the global length scale of the configuration.
In this respect, the magnetic null points are just a degenerate case of magnetic minima, whose associated QSLs are transformed into genuine separatrix surfaces.
More detailed studies are certainly needed in the future to quantify the relation between QSLs and magnetic minima. 
}}
%

%
%

\section{SUMMARY \label{s:sum}}

We have developed a general theory for the evolution of magnetic connections between boundary plasma elements in an arbitrary 3D magnetic configuration with an arbitrarily shaped boundary.
The theory assumes that at each instant the magnetic field is known everywhere throughout the volume domain including the boundary, where the tangential flow of plasma elements is also known.

The major quantity that is used for characterizing the magnetic connectivity and its evolution is the squashing factor $Q$ \citep{Titov1999a, Titov2007a}.
After being applied to the magnetic field-line mapping, the $Q$ factor allows one to identify in a given configuration all separatrix and quasi-separatrix surfaces, as well as their hybrids.
We have demonstrated first how the $Q$ factor is determined for any ideal evolution of the magnetic configuration in terms of the initial magnetic field and tangential boundary flows without reference to the magnetic field at other moments in time.
This result is itself of great importance for the further development of the theory of current-layer formations in magnetic configurations with a highly conducting plasma via its motion at the boundary.

In order to describe the nonideal evolution of magnetic connectivity, we have introduced the so-called ``slip mapping", which characterizes the change of magnetic connections between the boundary plasma elements due to violation of the frozen-in law in the plasma volume.
This mapping can be defined either forward or backward in time to give correspondingly slip-forth or slip-back mappings, which operate on the initial or final configurations, respectively, within a chosen time interval.
By calculating the squashing factor for these new mappings, we have defined new quantities called slip-forth and slip-back squashing factors, $Q_{\mathrm{sf}}$ and $Q_{\mathrm{sb}}$, respectively.
The large values of these quantities identify the reconnecting magnetic flux tubes, called reconnection fronts (RFs), at the initial and final time instances.
The boundary areas that are swept by RFs within a chosen time interval correspond to the footprints of reconnecting magnetic flux tubes.
In this way, the $Q_{\mathrm{sf}}$ factor determines in the initial configuration the to-be-reconnected magnetic flux tubes, while the $Q_{\mathrm{sb}}$ factor determines in the final configuration the reconnected magnetic flux tubes. 
The dependence of the properties of the RFs on tangential boundary flows is also analyzed.

{{\hl
The general results are illustrated with two specific examples based on evolving magnetic configurations.
The first of these was considered earlier by \citet{Hesse2005} in the framework of the general magnetic reconnection (GMR) theory \citep{Schindler1988, Hesse1988, Hesse1991, Hesse1993}.
Revisiting this example, we have demonstrated that our approach is consistent with GMR theory but complements it in several important aspects.
The second example describes a magnetic evolution that is driven by a thin current layer with a field-line voltage drop at a small patch inside it.
This example enables us to show that the localization of such a nonideal process in a small region may indeed cause the structural changes that are characteristic for reconnection rather than for magnetic diffusion.
The latter example, in particular, demonstrates the formation of magnetic flux tubes with narrow, ribbon-like footprints, which is a typical feature of reconnection process.

In comparison with GMR theory, our approach handles regions with small or large amounts of reconnected flux on an equal footing.
This is because the slip-squashing factors, in contrast to the voltage drop of GMR, are dimensionless geometrical quantities measuring only a relative spatial rate of magnetic slippage at the boundary.
For the same reason, the slip-squashing factors discriminate between regions of reconnection and simple diffusion much better than the voltage drop method.
Moreover, since the slip mappings are composed from the field-line mappings and both types of mappings are characterized by the same quantity, the analysis of reconnection based on our theory is intimately related to the analysis of magnetic structure.
As we have demonstrated here, the latter provides intriguing opportunities for analyzing magnetic reconnection in 3D configurations with an unprecedented level of detail.
The use of slip-squashing factors opens new perspectives for the comparing reconnecting flux tubes predicted by the theoretical models with the different morphological features observed in the solar flares and coronal mass ejections [e.g. brightening coronal loops, chromospheric kernels and ribbons, see the recent review by \citet{Demoulin2007}].

Our analysis of these examples also reveals the special role of hyperbolic and cusp minimum points in 3D magnetic field configurations.
The elemental magnetic flux tubes passing in the vicinity of such points may become extremely squashed, which implies the formation of the respective QSLs.
We have extended these examples by studying the field-line structure in the neighborhood of generic magnetic minima and confirmed this particular result.
}}
Hence, similar to magnetic null points, hyperbolic and cusp minima provide an important localization of plausible reconnection sites.
More detailed investigations in this direction promise to lead to even more interesting results.

It should be emphasized that, similar to GMR theory, our approach is applicable to the analysis of reconnection in magnetic configurations described not only by MHD, but also by any other plasma model.
The algorithm of calculating slip-squashing factors admits a straightforward implementation and it is highly parallelizable.
Our experience shows that, even without parallelization, the squashing factor may be computed for known magnetic field data with a modest demand on computing resources. 
Thus, the whole approach is practical enough to pursue the implementation of the proposed algorithms for calculating slip-squashing factors for the solar corona.
The solar physics group at SAIC is presently working on this project.


\acknowledgments

{\hl
We are grateful to the anonymous referee for valuable and inspiring discussions.
}
V.S.T. thanks Janvier Wijaya for his generous help with handling his nice application for plotting magnetic field lines.
V.S.T., T.G.F., and E.R.P., gratefully acknowledge the hospitality of the Isaac Newton Institute in Cambridge, UK, where the first formulation of the investigated problem has been appeared.
The contribution of V.S.T., Z.M., and J.R.L. was supported by NASA, the Center for Integrated Space Weather Modeling (an NSF Science and Technology Center), and by AFOSR contract FA9550-07-C-0044, T.G.F. by NSF grant ATM-0518218, and E.R.P. by the EU Solaire Network.






\clearpage


\end{document}